\documentclass[fleqn,usenatbib]{mnras}

\usepackage{newtxtext,newtxmath}
\usepackage{amsmath}
\usepackage{graphicx}
\usepackage{float}
\usepackage[T1]{fontenc}

\DeclareRobustCommand{\VAN}[3]{#2}
\let\VANthebibliography\thebibliography
\def\thebibliography{\DeclareRobustCommand{\VAN}[3]{##3}\VANthebibliography}

\usepackage{graphicx}	% Including figure files
\usepackage{amsmath}	% Advanced maths commands
\title[Globular Cluster X-ray Binaries in NGC 4261]{ The X-ray Point Source Population Hosted by Globular Clusters in the Elliptical Galaxy NGC 4261}

\author[S. Nair et al.]{
Sneha Nair,$^{1}$ $^{2} $\thanks{E-mail: sneha.nair2@mail.mcgill.ca}
Kristen C.  Dage,$^{1}$ $^{2}$ \thanks{E-mail: kristen.dage@mcgill.ca}
Daryl Haggard,$^{1}$ $^{2}$
Arunav Kundu, $^{3}$ $^{4}$
\newauthor
Richard M. Plotkin,$^{5}$ $^{6}$
Katherine L. Rhode,$^{7}$
Stephen E. Zepf $^{8}$
\\
$^{1}$ McGill Space Insitute,McGill University,3550 University Street,Montreal,QC H3A 2T8,Canada \\
$^{2}$Department of Physics,McGill University,3600 University Street,Montreal,QC H3A 2A7,Canada  \\
$^{3}$Eureka Scientific Inc., 2452 Delmer Street, Suite 100 Oakland, CA 94602, USA\\
$^{4}$ Department of Physics, Birla Institute of Technology \& Science, Pilani, K K Birla Goa Campus, NH17 B, Zuarinagar, Goa 403726, India\\
$^{5}$Department of Physics, University of Nevada, Reno, NV 89557, USA\\
$^{6}$Nevada Center for Astrophysics, University of Nevada, Las Vegas, NV 89154, USA \\
$^{7}$Indiana University Department of Astronomy, 727 East 3rd Street, Swain West 319, Bloomington, IN 47405, USA \\
$^{8}$Department  of  Physics  and  Astronomy,  Michigan  State  University,  East Lansing, MI 48824, USA
}

\date{Accepted XXX. Received YYY; in original form ZZZ}

\pubyear{2022}

\begin{document}
\label{firstpage}
\pagerange{\pageref{firstpage}--\pageref{lastpage}}
\maketitle

\begin{abstract}
Utilising archival \textit{Chandra X-ray Observatory} data and \textit{Hubble Space Telescope} globular cluster catalogues, we probe the time-domain properties of the low mass X-ray binary population in the elliptical galaxy NGC 4261. Of the 98 unique X-ray sources identified in this study, 62 sources are within the optical field of view and, of those, 33\% are aligned with an optical cluster counterpart. We find twenty X-ray sources coincident with globular clusters; two are previously discovered ultra-luminous X-ray sources (ULXs) and eighteen are low mass X-ray binaries (GCLMXBs) with $L_X < 10^{39}$ erg s$^{-1}$. ULXs are a heterogeneous class of extremely bright X-ray binaries ($L_X > 10^{39}$ erg s$^{-1}$) and ULXs located in globular clusters (GCULXs) and may be indicators of black holes. Identifying these unusually X-ray bright sources and measuring their optical properties can provide valuable constraints on the progenitors of gravitational wave sources. We compare observations of these sources to the twenty previously-studied GCULXs from five other early-type galaxies, and find that GCULXs in NGC 4261 are of similar colour and luminosity and do not significantly deviate from the rest of the sample in terms of distance from the galaxy centre or X-ray luminosity. Both the GCULX and low mass X-ray binary (GCLMXB) populations of NGC 4261 show long term variability; the former  may have implications for fast radio bursts originating in globular clusters and the latter will likely introduce additional scatter into the low mass end of GCLMXB X-ray luminosity functions.

\end{abstract}

%ULXs found in globular clusters (GCULXs) tend to be located in old and dense stellar systems. One proposed theory suggests that these GCULXs indicate the presence of an intermediate mass black hole, radiating at sub-Eddington rates. Another suggests that these might be stellar mass black holes radiating at super-Eddington rates. While the paucity of intermediate mass black hole candidates is well-known, these unusual sources are over represented in the sample of GCULXs. 

\begin{keywords}
NGC 4261: globular clusters:general – stars: black holes – X-rays: binaries 
\end{keywords}

%%%%%%%%%%%%%%%%%%%%%%%%%%%%%%%%%%%%%%%%%%%%%%%%%%

%%%%%%%%%%%%%%%%% BODY OF PAPER %%%%%%%%%%%%%%%%%%
%\sectionfont{\fontsize{12}{15}\selectfont}
\section{Introduction}
%DH: I restructured these first three paragraphs, but need a couple of refs to be added (originals still included below):
Ultraluminous X-ray sources (ULXs), or off-nuclear X-ray binaries with $L_x$ $>$ $10^{39}$ erg s$^{-1}$ in the 0.2-10 keV band (the Eddington limit for a 10 \(\textup{M}_\odot\) black hole), may either be indicators of some of the most extreme accretion physics in X-ray binaries, or represent a heterogeneous class of black holes \citep{fabbiano89}. Several ULXs are also known to exhibit evidence of a neutron star primary \citep[e.g.,][]{2014Natur.514..202B,2022MNRAS.511.5346S}. ULXs are most often found in active, young star forming regions of spiral galaxies \citep{swartz, kovlakas}, see also recent review by \citet{2023NewAR..9601672K}.  However, ULXs are also hosted in
globular clusters (GC), dense clusters of bound stars known to contain some of the oldest stars in their respective galaxies.  A total of 20 ULXs have been well-studied in GCs, including long-term investigations of their accretion signatures in both X-ray and optical observations. This has enabled studies of their optical properties and spatial distributions relative to host galaxy centres of the GCs \citep{2007Natur.445..183M, Shih, Irwin, Maccarone11, 2012ApJ...760..135R, Dage19a, Dage19b, Dage2020, Dage21}. 

The study of  X-ray binaries in globular clusters (down to limiting magnitudes of $\sim 10^{37}$ erg/s for early-type galaxies at the distances of Virgo and Fornax) offers a clean and unambiguous sample of low mass X-ray binaries (GCLMXBs) to probe the end-points of stellar evolution \citep[as a few examples, see][]{1995ApJ...439..687B, 2001ApJ...556..533S, 2005ApJ...631..511I,chiessantos2006,2006ApJ...647..276K, mk07, 2007ApJ...662..525K,2007ApJ...660.1246S, 2008ApJ...689..983H, 2013ApJ...764...98K,2017MNRAS.466.4021P,Lehmer20,2022arXiv220607192H,2022hxga.book..105G}. These works compare the presence of a GCLMXB to the optical properties and distribution of the host clusters, in particular linking optical colour/metallicity with the presence of an X-ray binary \citep[e.g.,][]{Kundu2003}. Of all the X-ray binaries, however, ULXs remain the most enigmatic due to their high X-ray luminosities. 

The broader-scale study of ULXs in globular clusters has offered clues to their nature, including evidence that globular cluster ULXs (GCULXs) are a heterogeneous class. The observed differences in X-ray and optical properties are due to both super-Eddington and near or sub-Eddington accretion \citep{Dage19a}, with the super-Eddington sources generating optical emission that may be linked to the shape of the X-ray spectrum.  \cite{Dage2020, Dage21} analysed the host cluster populations and found that while brighter clusters are more likely to host ULXs, there was no evidence for ULXs in the current sample to be preferentially hosted by more metal rich (redder) clusters. On the other hand, analysis by \cite{Kundu2003} shows that metal-rich clusters are three times more likely to host regular low mass X-ray binaries (GCLMXBs).

Since 2007, GCULXs have presented some of the first evidence for black holes in GCs \citep{2007Natur.445..183M}. Thus, ULXs in globular clusters can shed light on a longstanding debate in astronomy since theoretical work beginning with \cite{1969ApJ...158L.139S} suggested that black holes would be ejected from the host cluster. This question has increased in importance with the detection of gravitational wave events, as many more black hole-black hole binary mergers are being detected than were expected. Crucial to this work, the dynamical formation of black hole binaries in GCs is postulated as one of the primary formation channels \citep{abbott16}. 

Recent theoretical studies (e.g. \citealt{Morscher, rodriguez2016, giersz19, kremer19,2022arXiv220901564L}) have shown that GCs may retain more black holes than predicted by early models, and observational evidence for Galactic black hole candidates is rising (e.g., \citealt{strader12, 2013ApJ...777...69C, MillerJones15, giesers18, giesers19}). However, given that fewer than 200 GCs reside in the Milky Way and not all are equally likely to host BHs \citep{2020ApJ...898..162W}, it is important to turn to extragalactic GCs to identify a larger sample of black hole candidates in GCs.

 Observations outside of our Galaxy provide their own set of unique challenges due to large source distances, which reduce the overall data quality while increasing the sample size. For instance, elliptical galaxies such as NGC 4261 have extensive globular cluster systems.  \citet{Bonfini} identified more than 700 globular cluster candidates in NGC 4261, with a spread in their optical colour/metallicity distribution, and an X-ray binary population that can be characterised by suitably deep \textit{Chandra} observations.
 An early X-ray and optical study of the GC system in NGC 4261 by \cite{2005ApJ...634..272G} notes the presence of two ULXs in GCs and many GCLMXBs.  Since then, \textit{Chandra} has acquired a new deep observation of NGC 4261, which we have analysed to identify and study the GCLMXB system.

  In this paper, we present results from combining high-resolution data from the \textit{Chandra X-ray Observatory} with the \textit{Hubble Space Telescope} globular cluster catalogue of  \cite{Bonfini} to identify two GCULXs and 18 GCLMXB sources in NGC 4261. These sources are present in  both observations of the galaxy, and many have also been identified in the list of GCLMXBs published by \cite{2005ApJ...634..272G}. %The ground-based Isaac Newton Telescope images used by \cite{2005ApJ...634..272G} covers a wider field-of-view than \cite{Bonfini}, which accounts for any difference between the GCLMXBs.
 In Section \ref{sec:2}, we describe the data analysis associated with the X-ray and HST data as well as the selection of GCULXs and LXMBs. Section \ref{sec3} discusses a comparison of the GCULXs in NGC 4261 to the previously-identified sample of X-ray binaries in extra-galactic GCs. The implications of this study are discussed in Section \ref{sec4}.

\section{Data and Analysis}
\label{sec:2}
\begin{figure}
    \begin{center}
       \includegraphics[width=3in]{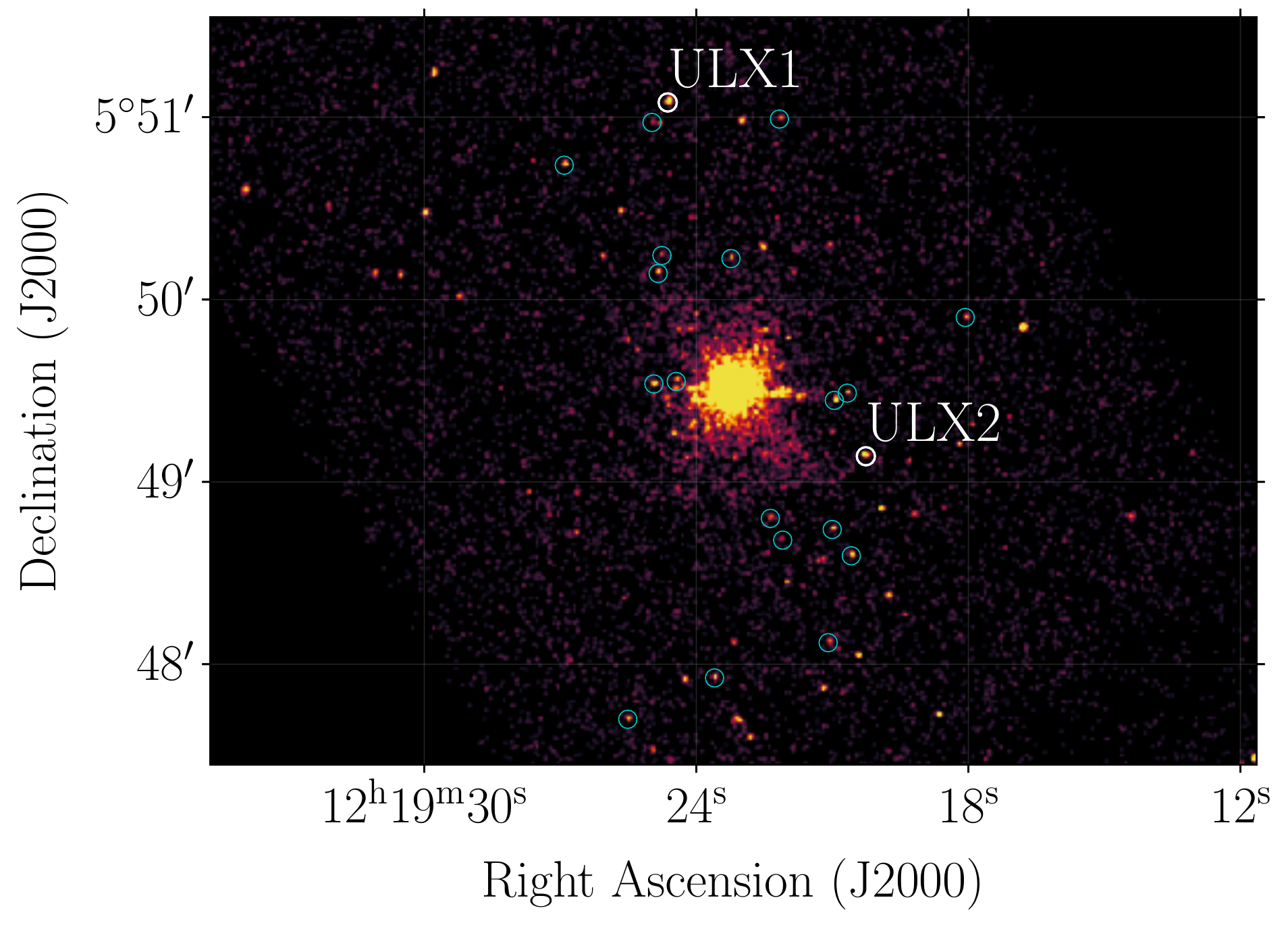} 
    \end{center}
    \caption{\textit{Chandra} image of NGC 4261 (ObsID 834) with regions corresponding to GCULXs (white, labeled regions) and GCLMXBs (blue regions) overlaid in the 0.5-8.0 keV band. We find eighteen GCLMXBs and two GCULXs.}
    \label{fig:img}
\end{figure}

The galaxy NGC 4261 was observed by the \textit{Chandra} X-ray Observatory with the Advanced CCD Imaging Spectrometer (ACIS-S) on May 6 2000 (ObsID 834, 34.40 ks, PI Birkinshaw) and on February 12 2008 (ObsID 9569, 100.94 ks, PI Zezas). The first of these observations was used to produce the image in Figure \ref{fig:img}, which shows the positions of the GCULXs and GCLMXBs identified in this work. ObsID 834 was taken in VFAINT mode while ObsID 9569 was obtained in FAINT mode. Throughout the analysis, we adopt a distance of 32 Mpc to NGC 4261 \citep{distance} and a 90\% confidence interval for all relevant uncertainties. We compare the X-ray point sources to the \cite{Bonfini} cluster catalogue in Section \ref{sec2.2}.

\subsection{X-ray Analysis}
\label{sec:ulxs} % used for referring to this section from elsewhere
  We analyze the \textit{Chandra} data using version 4.13 of the \textsc{ciao} \footnote{\url{https://cxc.cfa.harvard.edu/ciao/}} software package and \textsc{caldb} version 4.9.6, reprocessing both observations with \textsc{chandra\_repro}. We generate a list of X-ray point sources using the \textsc{wavdetect} algorithm, filtering each X-ray image  with an exposure map centered at an energy of 2.3 keV and wavelet scales of 1.0, 2.0, 4.0, 8.0 and 16.0. The enclosed fraction count (e.c.f.) parameter is set to 0.3 arcseconds. We set the significance threshold parameter at $10^{-6}$, corresponding to about one false detection per chip.  

  This produces a total of 63 X-ray point sources in ObsID 834, and 60 sources in ObsID 9569. We then implement \textsc{srcflux} using the source positions from \textsc{wavdetect} with background regions generated from the \textsc{roi}\footnote{\url{https://cxc.cfa.harvard.edu/ciao/ahelp/roi.html}} tool to measure the fluxes from the detected sources, assuming an absorbed power-law with a photon index of $\Gamma$= 1.7 and hydrogen column density\footnote{\url{https://cxc.harvard.edu/toolkit/colden.jsp}}, $N_H$=$1.58\times10^{20}$ cm$^{2}$. 

 ObsID 9569 was taken eight years after ObsID 834, and both cover the same central field. Therefore we are able to probe the time domain aspect of the GC X-ray sources. We estimate the limiting X-ray sensitivities in the 0.5-8.0 keV band for both observations, using \textsc{pimms} \footnote{\url{https://cxc.harvard.edu/toolkit/pimms.jsp}}. ObsID 834 has a limiting sensitivity around $3 \times 10^{37}$ erg s$^{-1}$ ( $2.5\times10^{-5}$ cts/sec) and the corresponding limit for ObsID 9569 is $\sim 1 \times 10^{37}$ erg s$^{-1}$ ( $6.9\times10^{-6}$ cts/sec). 
\subsection{Globular Cluster Counterpart Identification}
\label{sec2.2}
After ensuring that both the X-ray and optical data are aligned in the same astrometric frame, we use the globular cluster catalogue from \cite{Bonfini} to search for globular cluster counterparts to the detected X-ray sources in both ObsID 834 and ObsID 9569. We use \textsc{Topcat}\footnote{\url{http://www.star.bris.ac.uk/~mbt/topcat/}} to cross match between the X-ray point sources and optically selected GCs, with a matching radius of up to 1$"$. This results in a total of 20 X-ray sources that are co-spatial with a GC.  Of these, we classify two as GCULXs, which are also identified by \cite{2005ApJ...634..272G}. While our X-ray analysis of ObsID 834 recovers the same X-ray sources as \cite{2005ApJ...634..272G} in the central regions, we do not recover sources in the galaxy outskirts, as we only match to the HST globular cluster catalogue of \cite{Bonfini}. 

%\cite{2005ApJ...634..272G} used both HST observations as well as ground based observations from the Isaac Newton Telescope, which probed the outermost regions of the galaxy. 

\begin{table*}
	\centering
	\caption{X-ray and optical properties of GCULXs identified in NGC 4261. The X-ray count rates and fluxes were measured in the 0.5-8.0 keV band (in which \textit{Chandra} is most sensitive), and we used \textsc{pimms} to convert the brightest X-ray luminosity for the GC ULXs in the full band (0.2-10 keV). The peak luminosity of ULX1 in the full band is $2.94 \pm 0.53 \times 10^{39}$ erg s$^{-1}$, and ULX2 is $0.98 \pm 0.20\times 10^{39}$ erg s$^{-1}$.}
	\label{tab:ulx}
	\begin{tabular}{lcccccr} % four columns, alignment for each
		\hline
		\hline
		Name & RA & Dec &  L$_X$ (834) & L$_X$ (9569)   & $g$   &$g-z$\\
		&&& $10^{39}$erg s$^{-1}$ &$10^{39}$erg s$^{-1}$ &(AB Mag) & \\
		\hline
		GCULX1 &  12:19:24.631 &  +05:51:04.86 &$2.04^{+0.73}_{-0.61}$ &$2.25 \pm 0.41$ & 24.5&1.77 \\
		\\
		GCULX2 & 12:19:20.260 & +05:49:08.43 & $0.44^{+0.18}_{-0.10}$&$0.75 \pm 0.16$ &  24.4&1.45\\
		
	%	GCULX4 & 12:19:23.643 & +05:47:55.54 & 0.95 & 13.03 \pm 3.87&  22.9&1.45\\
		\hline
	\end{tabular}
\end{table*}
\section{Results and discussion}
\label{sec3}
\subsection{Globular Cluster X-ray Sources}
  We recover two of the \cite{2005ApJ...634..272G} GCULXs in NGC 4261 across two \textit{Chandra} observations (Table \ref{tab:ulx}). Hereafter, we refer to them as GCULX1 and GCULX2. GCULX1 is brighter than $2 \times 10^{39}$ erg s$^{-1}$ at both epochs while GCULX2 has a peak luminosity of $0.75 \times 10^{39}$ erg s$^{-1}$ in the 0.5-8.0 keV band in the second epoch. Conducting a band conversion for the 0.2-10.0 keV range with $\Gamma = 1.7$ brings the luminosity of GCULX2 to $10^{39}$ erg s$^{-1}$. Given that GCULX2 fits the nominal definition of a ULX in the 0.2-10.0 keV band, we include the source in the further analysis as a GCULX. We also detect eighteen X-ray sources coincident with GCs that have X-ray luminosity values on the order of $10^{38}$ erg s$^{-1}$ which are then classified as GCLMXBs.

\begin{figure}
    \begin{center}
       \includegraphics[width=3.5in]{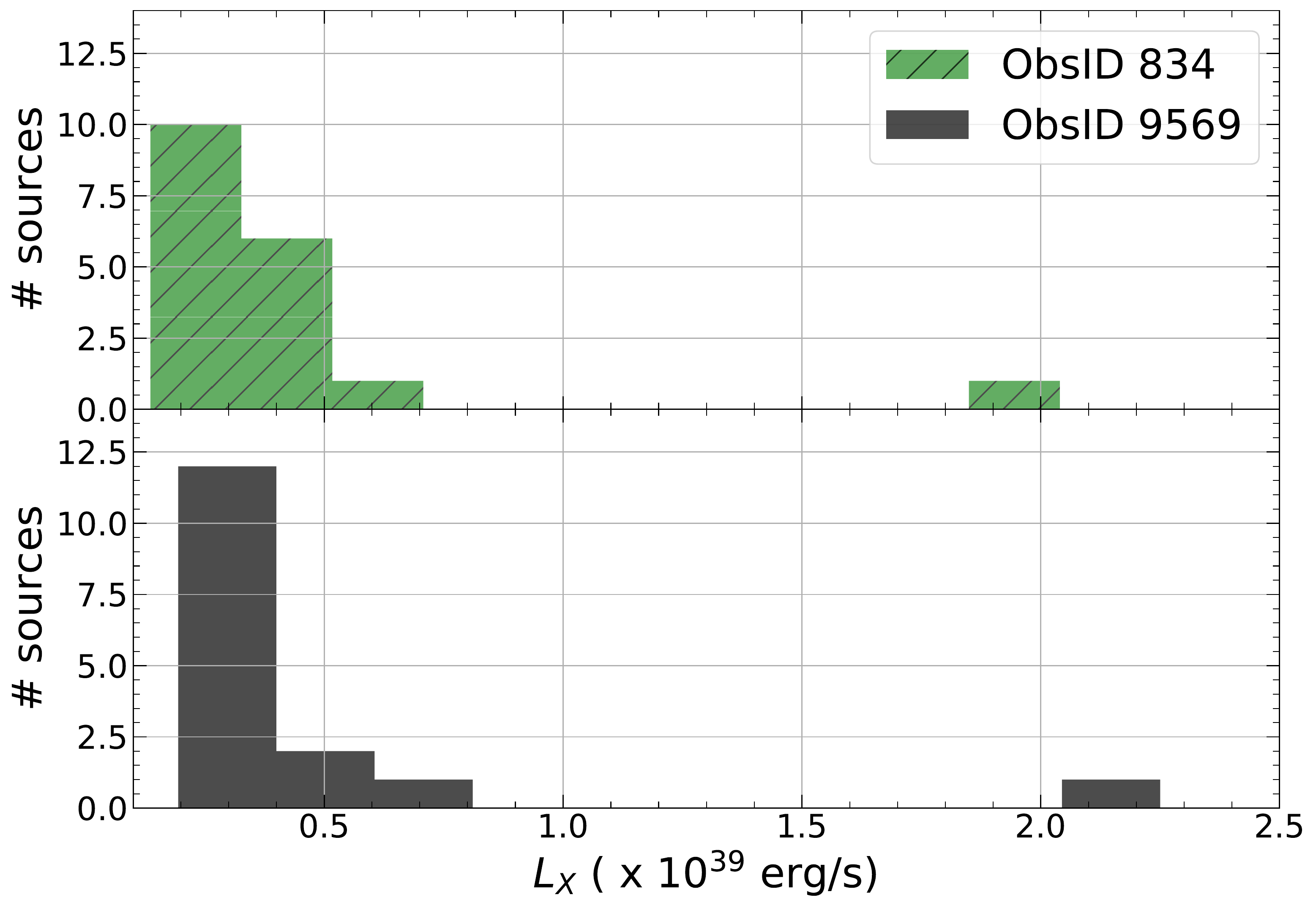} 
    \end{center}
    \caption{Histogram distribution of the 0.5-8.0 keV luminosities for the GCLMXB population in \textit{Chandra} ObsID 834 (06 May 2000) and ObsID 9569 (12 Feb 2008). %We note \textbf{some} variations in the luminosity distributions of GCLMXBs between the two epochs separated by nearly eight years.
    }
    \label{fig:yoyo}
\end{figure}

  \subsection{X-ray Variability}
  Given that both \textit{Chandra} observations have similar limiting sensitivities and both cover the HST field of view, we are able to track variability in the X-ray sources down to $\sim$ $3 \times 10^{37}$ erg s$^{-1}$ over the eight year time span. Both of the GCULXs appear in ObsID 834 and 9569. The X-ray luminosity of GCULX1 is self-consistent (within errors) across the two observations. By contrast, GCULX2 shows a factor of two increase in luminosity. Although our analysis would not have classified it as a ULX in ObsID 834, it became bright enough that that it exceeds the $10^{39}$ erg s$^{-1}$ Eddington limit in ObsID 9569. 
  
\begin{figure}
    \begin{center}
       \includegraphics[width=3.5in]{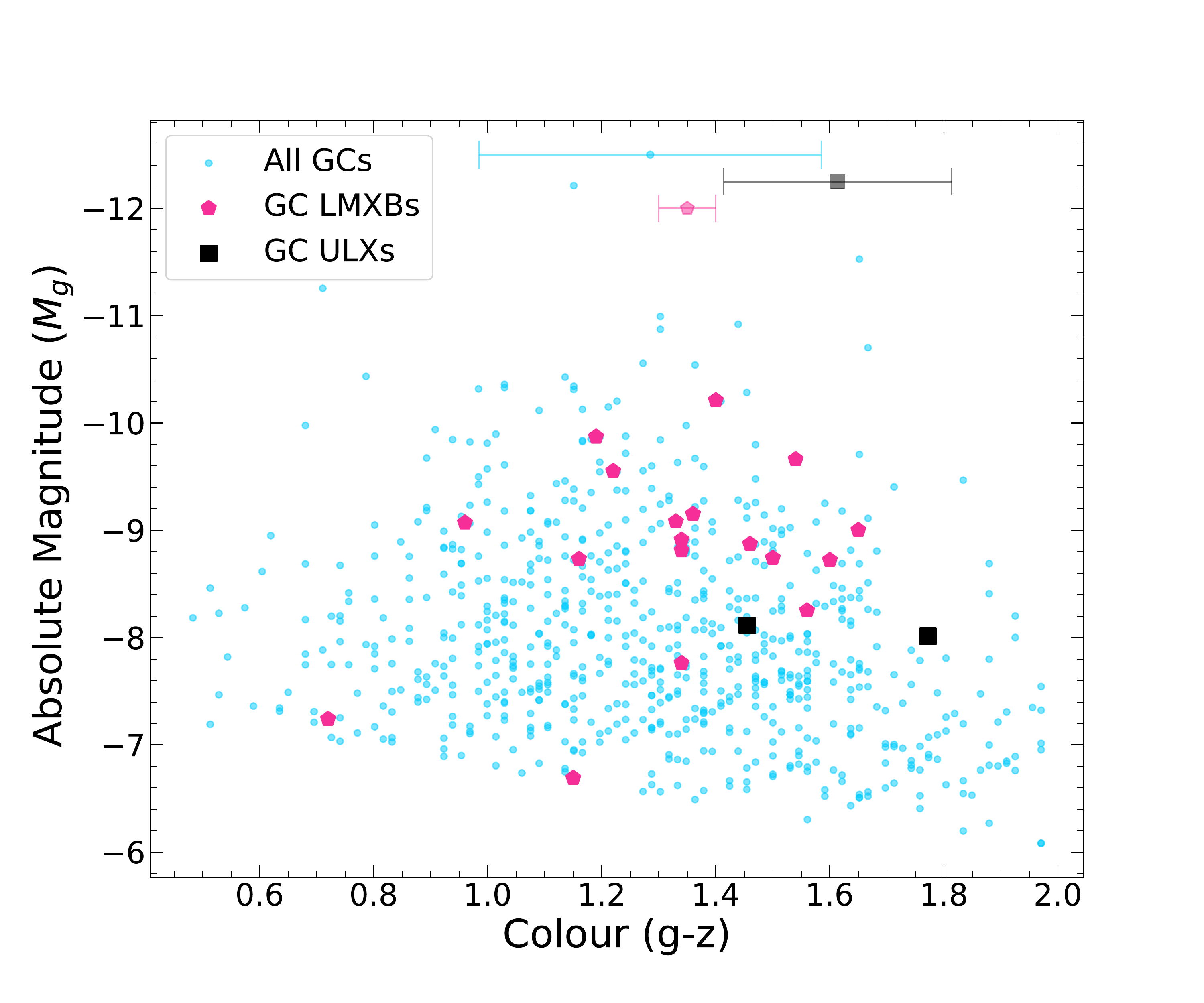} 
    \end{center}
    \caption{Colour-magnitude diagram of globular clusters (blue circles) in NGC 4261. ULX hosts are highlighted by purple squares and GCLMXBs hosts are denoted by pink pentagons. The mean colours of the GCLMXB hosts ($g-z$=$1.35 \pm 0.05$), GCULXs ($g-z$=$1.61 \pm 0.2$), as well as the overall GC population ($g-z$=$1.29 \pm 0.01$) are displayed in the error bars at an arbitrary y-value.} 
    \label{fig:blip}
\end{figure}
% GCULX2 sufficiently meets the threshold we have defined for a ULX, registering at $10.33  \times 10^{38}$ erg s$^{-1}$ at an energy band of 0.2-10 keV. We obtain slightly lower luminosities than \cite{2005ApJ...634..272G} as we use the the narrower energy band of 0.5-8.0 keV, which \textit{Chandra} is most sensitive to, and the same band that previous GCULX studies have been analyzed in. 
  
X-ray variability is an important diagnostic of the accretion behaviour of the systems, and also has important implications for fast radio bursts originating in globular clusters: studies such as \cite{2021ApJ...917...13S} favour ULXs as originators of fast radio bursts, but \cite{2022Natur.602..585K} note that X-ray limits rule out persistent X-ray sources for the globular cluster fast radio burst population. These studies do not, however, rule out X-ray sources that vary by an order of magnitude. Many GCULXs do show significant order of magnitude variability, in particular the source RZ2109 which shows an order of magnitude variability on the scale of hours, as well as over many years (e.g. \citealt{2007Natur.445..183M, Dage18}). Past studies of GCULXs have catalogued several different modes of unusual behaviour in GCULXs \citep{Shih, Dage19a, Dage2020, Dage21} and identified nine sources with no short/long-term variability (within uncertainties), eight more which vary over the course of many years but have no discernible variation within an observation (and have a 3 $\sigma$ upper limit on RMS variability, \citealt{Dage2020}), and three which show intra-observational variability (defined as the flux changing by at least a factor of two within an observation). One of these is a transient ULX which was observed by \textit{ROSAT} in the late 1990s, and by \textit{Chandra} until 2003, but never detected in subsequent observations \citep{Shih}.

 \begin{figure*}
     \renewcommand\thefigure{4} 
    \begin{center}
       \includegraphics[width=6in]{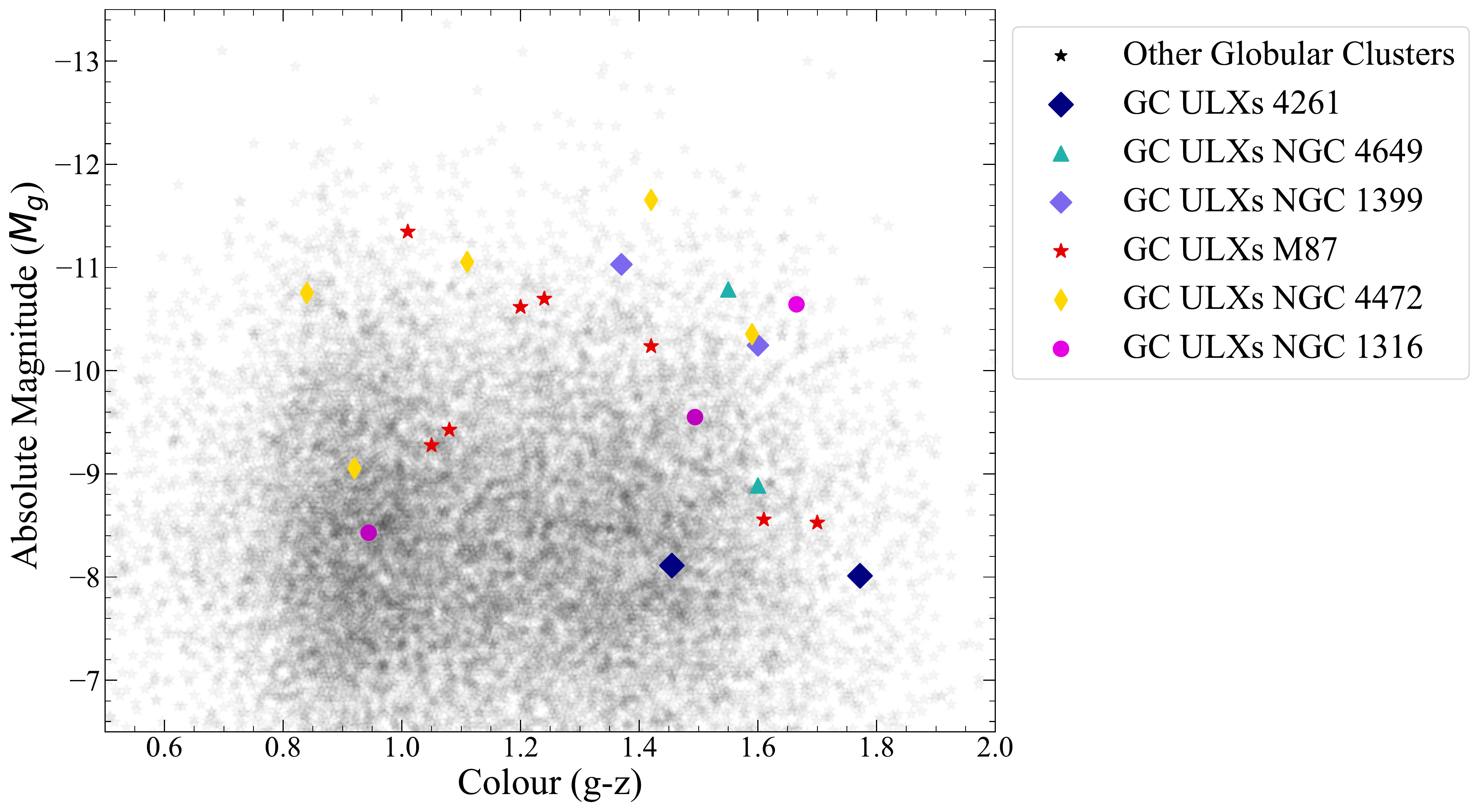} 
    \end{center}
    \caption{Colour-magnitude diagram of globular clusters in NGC 4261 (blue squares), with the grey points showing globular clusters from NGC 1316, NGC 1399, NGC 4472, NGC 4649 and M87 \citep{Dage19a, Dage2020, Dage21}. GCULXs from the aforementioned galaxies are highlighted by coloured square markers. The magnitudes and colors of GCULXs in NGC 4261 are similar to those of observed in other ellipticals. Although formally the NGC 4261 GCULX hosts appear to be less luminous and redder than counterparts in other galaxies these differences are not statistically robust as explained in Section 3.3. }
    \label{fig:blipblop}
\end{figure*}
Based on our analysis, NGC 4261 GCULX1 does not appear to vary in the observations at hand, but GCULX2 shows long-term variability. We did not detect significant intra-observational variability in either source, and used \textsc{lcstats} to place 3 $\sigma$ RMS upper limits of $\sim 30\%$ for the two ULXs in ObsID 834 and $\sim 60\%$ in ObsID 9569.  Although these sources are quite rare, the variability paradigms suggest different accretion physics and binary properties across the entire sample of GCULXs studied to date.

\begin{figure}
    \renewcommand\thefigure{5} 
    \centering
    \includegraphics[width=3in]{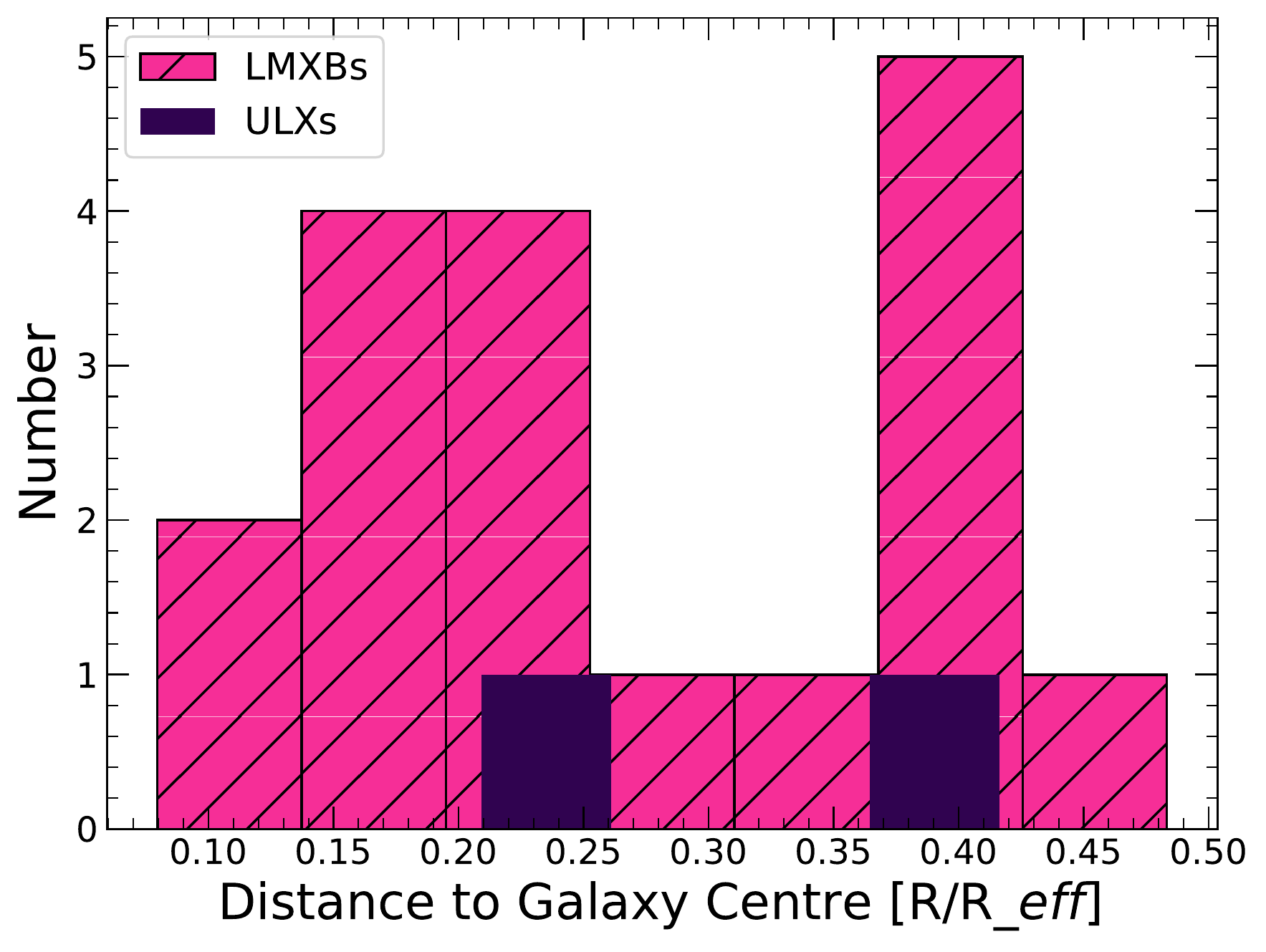}
    \caption{Spatial distribution of NGC 4261's GCULXs and GCLMXBs. Due to HST's smaller field of view, we find these sources relatively close to the centre of NGC 4261. }
    \label{fig:bootss}
\end{figure}

\begin{table*}
	\centering
	\renewcommand\thetable{2} 
	\caption{X-ray and optical properties of low mass X-ray binaries identified in NGC 4261 with their positions indicated. Sources that were identified in the later observation (ObsID 9569) are highlighted with an asterisk. Some X-ray sources were cross-matched to two possible GCs; in those cases, both optical values are presented. X-ray luminosities are in the 0.5-8.0 keV band. }
	\label{tab:lmxb}
	\begin{tabular}{lcccccr} % four columns, alignment for each
		\hline
		\hline
		Name & RA & Dec &  L$_X$ (834) & L$_X$ (9569)  & $g$   &$g-z$\\
		&&& $10^{38}$erg s$^{-1}$& $10^{38}$erg s$^{-1}$ &(AB Mag) & \\
		\hline
		\vspace{0.09cm}
		GCLMXB1 & 12:19:25.512 & +05:47:41.86 & $3.2^{+1.4}_{-1.1}$ & $3.0^{+1.1}_{-1.0}$  & 23.00 & 1.22\\
	%	&&&&&&&\\
	\vspace{0.09cm}
		GCLMXB2 & 12:19:23.598 & +05:47:55.48 & $5.3^{+2.9}_{-2.2}$ & - & 23.40/23.98&1.36/1.16\\
	%	&&&&&&&\\
	\vspace{0.09cm}
		GCLMXB3 & 12:19:21.092 & +05:48:07.09  & $2.0^{+1.5}_{-1.1}$ & -  & 25.31 & 0.72\\
	%	&&&&&&&\\
	\vspace{0.09cm}
		GCLMXB4 & 12:19:20.582 & +05:48:35.67 & $4.9^{+2.5}_{-1.9}$ &- & 23.74& 1.34\\
	%	&&&&&&&\\
	\vspace{0.09cm}
		GCLMXB5& 12:19:22.096 & +05:48:40.83 & $1.4^{+1.2}_{-0.8}$ & - &25.07/24.79& 1.86/1.34\\ %double match
		%&&&&&&&\\
		\vspace{0.09cm}
		GCLMXB6 & 12:19:21.003 & +05:48:44.33 & $2.8^{+1.5}_{-1.1}$ & $3.1^{+1.0}_{-0.9}$& 23.68 &1.46\\
	%	&&&&&&&\\
		\vspace{0.09cm}
		GCLMXB7 & 12:19:22.367 & +05:48:47.98 &  $3.5^{+2.0}_{-1.5}$ & - &23.82 &1.16 \\
	%	&&&&&&&\\
	\vspace{0.09cm}
		GCLMXB8 & 12:19:20.956  & +05:49:26.81 & $3.4^{+1.6}_{-1.2}$ & $3.2^{+1.0}_{-0.8}$   & 23.81  &1.50\\
	%	&&&&&&&\\
	\vspace{0.09cm}
		GCLMXB9 & 12:19:20.670 & +05:49:29.20  & $1.5^{+1.2}_{-0.8}$ &$3.5^{+1.0}_{-0.9}$ &23.55 & 1.65\\
	%	&&&&&&&\\
	\vspace{0.09cm}
		
		GCLMXB10& 12:19:24.936 & +05:49:32.23 & $3.1^{+1.7}_{-1.3}$ & $4.4^{+1.3}_{-1.1}$  &23.64 &1.34\\
	%	&&&&&&&\\
	\vspace{0.09cm}
		GCLMXB11& 12:19:24.444 & +05:49:33.04 & $2.0^{+1.3}_{-0.9}$&- &24.18/23.48 &1.65/0.96\\ 
	%	&&&&&&&\\
	\vspace{0.09cm}
		GCLMXB12 & 12:19:18.068 & +05:49:54.10 & $1.4^{+1.2}_{-0.8}$& $2.7^{+0.8}_{-1.1}$ &23.83 &1.60 \\
	%	&&&&&&&\\
	\vspace{0.09cm}
		GCLMXB13& 12:19:24.842 & +05:50:08.59 & $3.9^{+3.0}_{-2.2}$ &$3.7^{+1.2}_{-1.1}$  & 22.68 & 1.19\\
	%	&&&&&&&\\
	\vspace{0.09cm}
		GCLMXB14 & 12:19:23.236 & +05:50:13.49& $1.6^{+1.2}_{-0.9}$ &$1.9^{+1.0}_{-1.1}$ &  24.30 &1.56\\
	%	&&&&&&&\\
	\vspace{0.09cm}
		GCLMXB15 & 12:19:26.914 & +05:50:44.21 & $4.7^{+2.7}_{-1.6}$& -& 22.34&1.40 \\
	%	&&&&&&&\\
	\vspace{0.09cm}
	    GCLMXB16 & 12:19:22.167 &+05:50:59.41  & $2.3^{+1.6}_{-1.1}$ &$2.1^{+0.7}_{-1.1}$ & 25.86&1.15\\
	%	&&&&&&&\\
	\vspace{0.09cm}
		GCLMXB17* & 12:19:24.759 & +05:50:14.50 & - & $2.6^{+1.6}_{-1.2}$  & 23.47  & 1.33\\
		%&&&&&&&\\
		\vspace{0.09cm}
		GCLMXB18* & 12:19:24.978 &+05:50:58.28 & - &$3.6^{+1.2}_{-1.0}$ &   22.89 & 1.54 \\
		
		\hline
	\end{tabular}
\end{table*}   
   
When we examine the rest of the GCLMXBs, we find two in ObsID 9569 that were not detected, although both observations covered the same field, and we find four GCLMXBs that ``shut off'' in ObsID 9569. The rest of the LXMBs do not differ significantly within their measurement uncertainties (Table \ref{tab:lmxb}).

GCULX variability has important implications in other fields: for instance, the X-ray binary population in globular clusters can be used to characterise the low mass X-ray binary population \citep{add,2010ApJ...724..559L,Lehmer20}, which in turn has implications for classifying populations of black holes and AGN \citep{Lemons15}. GCULXs pin down the end points of these functions, and their variability may cause additional scatter in the scaling relations beyond normal statistics, as suggested by the histogram of the GCLMXB X-ray luminosities across observations (Figure \ref{fig:yoyo}). Indeed, in the first observation of NGC 4261, GCULX2 barely meets the threshold for a ULX, whereas eight years later, it reached a luminosity where it is securely classified as a ULX. This suggests that in addition to finding GCLMXB populations in many galaxies to improve statistics, we need to consider the time variability of these X-ray sources when trying to accurately determine the GCLMXB X-ray luminosity function.

%This, and other time domain aspects of GCULXs suggests that in addition to searching and combining GCLMXB populations in many galaxies, the time domain aspect may also play a role in the GCLMXB X-ray luminosity function.

%\subsection{X-ray Luminosity Functions}
%% Rough Notes on what I might write
%To analyze the behaviour of the GCLMXBs changing through the observations we create three XLF. Two for each observation and one with the total 20 X-ray sources that have been matched to a GC . Clear differences in the observation that is 8 years later, we can clearly see that the power law fit drops off at a steeper rate. Report the slopes that you get from the fit. It implies that in this observation the number of sources approaching the luminosity or exceeding drops off much quicker than in the earlier observation. Notably, this is interesting because even though we account for the fact that Chandra collects dust in that time the limiting luminosities for both observation  pretty much stay the same . They are in that sense comparable so the fact that in the second observation our ULXs get brighter is strange. Make some comparison 
%Ask Kristen:About maybe doing broken power law model or why they use it in some papers. 

 \begin{figure}
    \renewcommand\thefigure{6} 
    \begin{center}
       \includegraphics[width=3in]{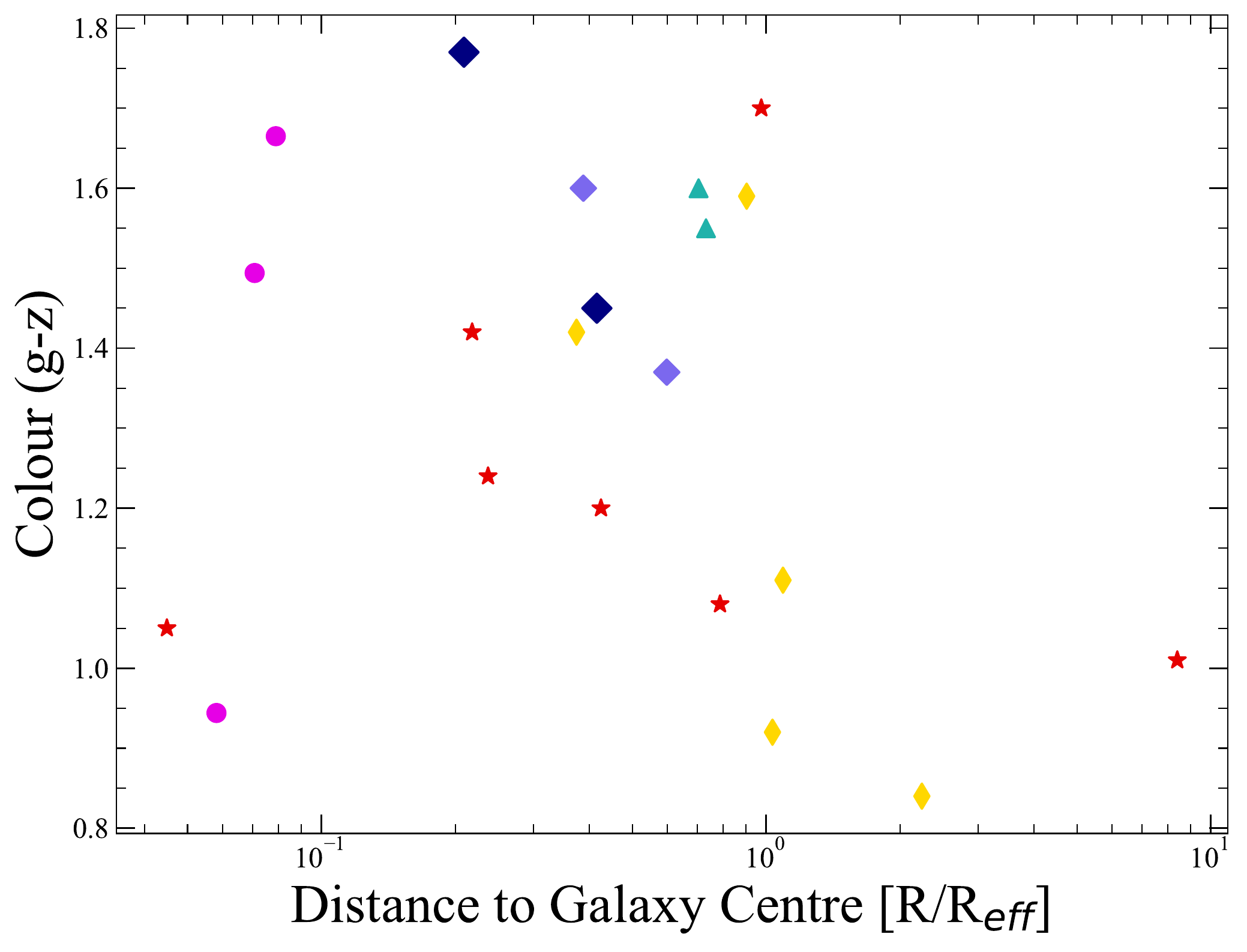} 
    \end{center}
    \caption{Optical colours of known GCULXs compared to their distance to the centre of the galaxy in terms of effective radius, including NGC 4472, NGC 1316, NGC 4649, NGC 1399 and M87 \citep{Dage19a, Dage2020, Dage21}, alongside NGC 4261 (this work). Note that the symbols follow the same legend as in Figure \ref{fig:blipblop}.  We note that there are more GCULXs detected closer in to the galaxy centres; this may be due to an observational bias, as the outer regions of external galaxies at these distances are often unobserved by \textit{HST}. }
\label{fig:rojo2}
\end{figure}

\begin{table}
\renewcommand\thetable{3} 
\caption{Mean optical colour ($g-z$) for the \citet{Bonfini} GC catalogue,  GCLMXBs, and GCULXs.}
\label{tab:mean}
\begin{center}
\begin{tabular}{|l|l|c|}
\hline
Grouping       & Mean Colour   \\ \hline
All GCs     & 1.29  $\pm$ 0.01  \\ 
GCLMXB        & 1.35 $\pm$ 0.05 \\ 
GCULXs & 1.61          $\pm$ 0.2  \\ \hline
%GCULXs (all)  & 1.34        & 0.28               & 57                       \\ \hline
\end{tabular}
\label{mean}
\end{center}
\end{table}

\subsection{Host Cluster Colour and Magnitude}
\label{gurg}

 We use the relations of \cite{Jester05} and \cite{Peacock2010} to convert the V and I optical magnitudes in \cite{Bonfini}  to $g$ and $z$ magnitudes\footnote{($g-z$) = 1.518$\times$(V-I)-0.443 and $g$ =V+0.39$\times$($g-z$)-0.08}, since the majority of the previously studied GCULXs have colours presented in the $g-z$ system. The colour-magnitude diagram of the globular clusters in NGC 4261 is presented in Figure \ref{fig:blip}. The mean GCLMXB colour is 1.35 $\pm$ 0.05, and the mean GCULX host cluster colour is 1.61 $\pm 0.2$  (see Table \ref{tab:mean}).  We note that most of the GCLMXB-hosting globular clusters in NGC 4261 (Table \ref{tab:xray1}) have redder optical colours, which is similar to the trend identified by \cite{Kundu2003}, who found a GCLMXB mean colour of $g-z$ = 1.22 \footnote{ \cite{Kundu2003} reports <V-I>=1.10.}, with a three to one preference for redder GCs to host GCLMXBs.  \cite{Dage21} found that the host clusters of previously identified GCULXs spanned a wide range of optical colours. The mean colours of the entire cluster sample are displayed in Table \ref{tab:mean}.

 An Anderson-Darling test comparing the optical properties of the 20 clusters in NGC 4261 which host GCLMXBs to the rest of the GCs gives a statistic of 0.008 and a significance level of 0.25 for optical colour, and a statistic of 7.8 and significance level of 0.001 for magnitude. This further suggests that the clusters which host GCLMXBs are distinct populations in magnitude at high significance ($>3 \sigma$).
 
 While Figure 3 appears to suggest that$-$ unlike previous studies of GCULXs$-$ NGC 4261's ULXs do not appear to be preferentially hosted by brighter clusters, but do have metal-rich hosts these correlations should be considered with an abundance of caution due to the pecularities of the Hubble Space Telescope data set used in the optical analysis of \cite{Bonfini}. The analysis is based on relatively shallow 800s snapshot data of each field for this distant galaxy at 32 Mpc. Furthermore, the HST observations were obtained in late 2007 and early 2008 when the WFPC2 camera aboard the HST was near its end of life and suffered from significant charge transfer efficiency loss \citep{Dolphin_2009}, which adds to the uncertainties in photometric parameters.  Figure \ref{fig:blipblop} shows the colour magnitude diagram of the NGC 4261 ULX hosts compared to the other GCULXs as well as the globular cluster populations in all of the six previously studied galaxies. %In addition, it is evident that the ULX host clusters in NGC 4261 are more metal rich than the other ULX hosting clusters. 
 This suggests that there is indeed a large variance in the small population of GCULXs and their host clusters.
 
\subsection{Spatial Distribution of ULX Hosting Clusters}
We find that the GCULXs in NGC 4261 are located close to the host galaxy centre (Figure \ref{fig:bootss}).  Of the previously studied GCULXs, 17 are located within two effective radii of their respective host galaxy centres, and only three sources are more distant. We compare the optical colour of the host clusters to the distance from galaxy centre in units of the effective radius of the galaxy, (as measured by \citealt{2013MNRAS.432.1862C}), and show these results in Figure \ref{fig:rojo2}.

 The clusters that host ULXs span a wide range of optical colours, from very blue (metal-poor) to extremely very red (metal-rich), and there does not appear to be a significant trend between optical colour and spatial distance. However, we note that the clusters that are the most distant from their galaxy centres are also very blue. This is broadly consistent with what is expected for globular cluster populations of giant galaxies; i.e, blue globular clusters tend to have more extended radial distributions compared to redder populations \citep{brodie}.

%%INserting table here for now 

\section{Summary and conclusions}
\label{sec4}
We analyse two archival \textit{Chandra} X-ray observations of the elliptical galaxy NGC 4261. These have similar central fields of view and limiting sensitivities, but are separated by eight years. We search for GC counterparts to the X-ray point sources by using the cluster catalogue from \cite{Bonfini}. Our specific results include:

\begin{itemize}
\item We recover two ULX sources identified in GCs by \cite{2005ApJ...634..272G} and identify a total of 18 more X-ray sources with $L_X < 10^{39}$ erg s$^{-1}$ that are coincident with the positions of the globular cluster in the \cite{Bonfini} optical study (across the two observations).  
    \item The time domain properties of GCULX1 and GCULX2 are quite different from each other: GCULX1 remains consistently bright across the two observations, but GCULX2's X-ray luminosity is doubled in the second observation, thereby changing its classification.
    \item The GCLMXB population overall shows some variability, with four sources detected in the first observation but not in the second observation and two new sources detected in the second observation. Both observations have similar detection thresholds, which suggests a variable source population.
   \item The GCs in NGC 4261 that host GCLMXBs are more metal rich and luminous than the general population. While the two GCULXs formally appear to populate the redder and less optically luminous end of the GCLMXB sample this is not statistically significant because of small number statistics and the limitations of the optical data.  
\item  The optical luminosities (masses) and colors (metallicities) of the GCULXs in NGC 4261 are similar to the twenty previously studied candidates in five other early type galaxies. 

    \item We find that, while there is a wide spread in the optical colours of GCs that host ULXs, the ones most offset from the galaxy center are bluer, although there is not yet a large enough sample to test this.

%We find that the GCs in NGC 4261 that host ULXs are more metal-rich compared to the general population, \textbf{and} are hosted by fainter clusters, \textbf{albeit for a small sample of only two ULX sources}. \textbf{ Despite the two ULXs in NGC 4261 being somewhat redder/more metal-rich than many of the globular clusters that host ULXs in other galaxies (the median optical \textit{g-z} colour of the entire sample is 1.34 with a standard deviation of 0.28),  there is not yet strong evidence that GCULXs are preferentially situated in more metal rich environments. Again we caution that this is still a small sample size.}

%    \item Like most of the previously studied GCULXs, the GCULXs in NGC 4261 tend to reside close to the host galaxy's centre.%Follow-up studies that have more complete coverage of the outer regions of elliptical galaxies would be useful and allow this result to be examined further.

\end{itemize}

The X-ray sources identified in this study represent an important increase to the number of GCULXs, and thus the number of black holes in extragalactic GCs. They thereby provide additional opportunities to study objects that represent one of the primary formation channels for gravitational wave progenitors. Understanding the nature of strongly variable ultraluminous X-ray sources may also provide constraints on the connection between ultraluminous X-ray sources and fast radio bursts \citep{2021ApJ...917...13S}. Studies of GCLMXBs in extragalactic globular clusters such as those by \cite{Peacock17} and \cite{Lehmer20} identify a large number of X-ray sources in GCs below $10^{39}$ erg s$^{-1}$, but exceptionally few above this limit. In our own Galaxy, the cluster with the brightest integrated X-ray luminosity is M15, which reaches only a few $\times 10^{38}$ erg s$^{-1}$. Although the sample of well-studied GCULXs is still small, the large variation in behaviours points towards diversity in the make-up of the binary systems, and the need for identification of even more of these extreme systems in order to construct a more complete picture of their physical nature. 

\section*{Acknowledgements}
We thank the anonymous referee for the helpful feedback which greatly improved the paper. SN, KCD and DH acknowledge funding from the Natural Sciences and Engineering Research Council of Canada (NSERC), and the Canada Research Chairs (CRC) program. KCD acknowledges fellowship funding from the McGill Space Institute and from Fonds de Recherche du Qu\'ebec $-$ Nature et Technologies, Bourses de recherche postdoctorale B3X no. 319864. DH thanks the Institut de Plan{\'e}tologie et d'Astrophysique de Grenoble (IPAG) for supporting an extended visit during which this work was completed. SEZ acknowledges support from grant GO9-20080X. This work was performed in part at Aspen Center for Physics, which is supported by National Science Foundation grant PHY-1607611.

%%%%%%%%%%%%%%%%%%%%%%%%%%%%%%%%%%%%%%%%%%%%%%%%%%
\section*{Data Availability}
The \textit{Chandra} observations are publicly available at \url{https://cda.harvard.edu/chaser/}.


\bibliographystyle{mnras}
\bibliography{NGC4261} % if your bibtex file is called example.bib

\begin{thebibliography}{}
\makeatletter
\relax
\def\mn@urlcharsother{\let\do\@makeother \do\$\do\&\do\#\do\^\do\_\do\%\do\~}
\def\mn@doi{\begingroup\mn@urlcharsother \@ifnextchar [ {\mn@doi@}
  {\mn@doi@[]}}
\def\mn@doi@[#1]#2{\def\@tempa{#1}\ifx\@tempa\@empty \href
  {http://dx.doi.org/#2} {doi:#2}\else \href {http://dx.doi.org/#2} {#1}\fi
  \endgroup}
\def\mn@eprint#1#2{\mn@eprint@#1:#2::\@nil}
\def\mn@eprint@arXiv#1{\href {http://arxiv.org/abs/#1} {{\tt arXiv:#1}}}
\def\mn@eprint@dblp#1{\href {http://dblp.uni-trier.de/rec/bibtex/#1.xml}
  {dblp:#1}}
\def\mn@eprint@#1:#2:#3:#4\@nil{\def\@tempa {#1}\def\@tempb {#2}\def\@tempc
  {#3}\ifx \@tempc \@empty \let \@tempc \@tempb \let \@tempb \@tempa \fi \ifx
  \@tempb \@empty \def\@tempb {arXiv}\fi \@ifundefined
  {mn@eprint@\@tempb}{\@tempb:\@tempc}{\expandafter \expandafter \csname
  mn@eprint@\@tempb\endcsname \expandafter{\@tempc}}}

\bibitem[\protect\citeauthoryear{{Abbott} et~al.,}{{Abbott}
  et~al.}{2016}]{abbott16}
{Abbott} B.~P.,  et~al., 2016, \mn@doi [\apjl] {10.3847/2041-8205/833/1/L1},
  \href {https://ui.adsabs.harvard.edu/abs/2016ApJ...833L...1A} {833, L1}

\bibitem[\protect\citeauthoryear{{Bachetti} et~al.,}{{Bachetti}
  et~al.}{2014}]{2014Natur.514..202B}
{Bachetti} M.,  et~al., 2014, \mn@doi [\nat] {10.1038/nature13791}, \href
  {https://ui.adsabs.harvard.edu/abs/2014Natur.514..202B} {514, 202}

\bibitem[\protect\citeauthoryear{{Bellazzini}, {Pasquali}, {Federici},
  {Ferraro}  \& {Pecci}}{{Bellazzini} et~al.}{1995}]{1995ApJ...439..687B}
{Bellazzini} M.,  {Pasquali} A.,  {Federici} L.,  {Ferraro} F.~R.,   {Pecci}
  F.~F.,  1995, \mn@doi [\apj] {10.1086/175208}, \href
  {https://ui.adsabs.harvard.edu/abs/1995ApJ...439..687B} {439, 687}

\bibitem[\protect\citeauthoryear{{Bonfini}, {Zezas}, {Birkinshaw}, {Worrall},
  {Fabbiano}, {O'Sullivan}, {Trinchieri}  \& {Wolter}}{{Bonfini}
  et~al.}{2012}]{Bonfini}
{Bonfini} P.,  {Zezas} A.,  {Birkinshaw} M.,  {Worrall} D.~M.,  {Fabbiano} G.,
  {O'Sullivan} E.,  {Trinchieri} G.,   {Wolter} A.,  2012, \mn@doi [\mnras]
  {10.1111/j.1365-2966.2012.20514.x}, \href
  {https://ui.adsabs.harvard.edu/abs/2012MNRAS.421.2872B} {421, 2872}

\bibitem[\protect\citeauthoryear{{Brodie} \& {Strader}}{{Brodie} \&
  {Strader}}{2006}]{brodie}
{Brodie} J.~P.,  {Strader} J.,  2006, \mn@doi [\araa]
  {10.1146/annurev.astro.44.051905.092441}, \href
  {https://ui.adsabs.harvard.edu/abs/2006ARA&A..44..193B} {44, 193}

\bibitem[\protect\citeauthoryear{{Cappellari} et~al.,}{{Cappellari}
  et~al.}{2013}]{2013MNRAS.432.1862C}
{Cappellari} M.,  et~al., 2013, \mn@doi [\mnras] {10.1093/mnras/stt644}, \href
  {https://ui.adsabs.harvard.edu/abs/2013MNRAS.432.1862C} {432, 1862}

\bibitem[\protect\citeauthoryear{{Chies-Santos}, {Pastoriza}, {Santiago}  \&
  {Forbes}}{{Chies-Santos} et~al.}{2006}]{chiessantos2006}
{Chies-Santos} A.~L.,  {Pastoriza} M.~G.,  {Santiago} B.~X.,   {Forbes} D.~A.,
  2006, \mn@doi [\aap] {10.1051/0004-6361:20054212}, \href
  {https://ui.adsabs.harvard.edu/abs/2006A&A...455..453C} {455, 453}

\bibitem[\protect\citeauthoryear{{Chomiuk}, {Strader}, {Maccarone},
  {Miller-Jones}, {Heinke}, {Noyola}, {Seth}  \& {Ransom}}{{Chomiuk}
  et~al.}{2013}]{2013ApJ...777...69C}
{Chomiuk} L.,  {Strader} J.,  {Maccarone} T.~J.,  {Miller-Jones} J. C.~A.,
  {Heinke} C.,  {Noyola} E.,  {Seth} A.~C.,   {Ransom} S.,  2013, \mn@doi
  [\apj] {10.1088/0004-637X/777/1/69}, \href
  {https://ui.adsabs.harvard.edu/abs/2013ApJ...777...69C} {777, 69}

\bibitem[\protect\citeauthoryear{{Dage}, {Zepf}, {Bahramian}, {Kundu},
  {Maccarone}  \& {Peacock}}{{Dage} et~al.}{2018}]{Dage18}
{Dage} K.~C.,  {Zepf} S.~E.,  {Bahramian} A.,  {Kundu} A.,  {Maccarone} T.~J.,
   {Peacock} M.~B.,  2018, \mn@doi [\apj] {10.3847/1538-4357/aacb2b}, \href
  {https://ui.adsabs.harvard.edu/abs/2018ApJ...862..108D} {862, 108}

\bibitem[\protect\citeauthoryear{{Dage}, {Zepf}, {Peacock}, {Bahramian},
  {Noroozi}, {Kundu}  \& {Maccarone}}{{Dage} et~al.}{2019a}]{Dage19a}
{Dage} K.~C.,  {Zepf} S.~E.,  {Peacock} M.~B.,  {Bahramian} A.,  {Noroozi} O.,
  {Kundu} A.,   {Maccarone} T.~J.,  2019a, \mn@doi [\mnras]
  {10.1093/mnras/stz479}, \href
  {https://ui.adsabs.harvard.edu/abs/2019MNRAS.485.1694D} {485, 1694}

\bibitem[\protect\citeauthoryear{{Dage} et~al.,}{{Dage}
  et~al.}{2019b}]{Dage19b}
{Dage} K.~C.,  et~al., 2019b, \mn@doi [\mnras] {10.1093/mnras/stz2514}, \href
  {https://ui.adsabs.harvard.edu/abs/2019MNRAS.489.4783D} {489, 4783}

\bibitem[\protect\citeauthoryear{{Dage}, {Zepf}, {Thygesen}, {Bahramian},
  {Kundu}, {Maccarone}, {Peacock}  \& {Strader}}{{Dage}
  et~al.}{2020}]{Dage2020}
{Dage} K.~C.,  {Zepf} S.~E.,  {Thygesen} E.,  {Bahramian} A.,  {Kundu} A.,
  {Maccarone} T.~J.,  {Peacock} M.~B.,   {Strader} J.,  2020, \mn@doi [\mnras]
  {10.1093/mnras/staa1963}, \href
  {https://ui.adsabs.harvard.edu/abs/2020MNRAS.497..596D} {497, 596}

\bibitem[\protect\citeauthoryear{{Dage} et~al.,}{{Dage} et~al.}{2021}]{Dage21}
{Dage} K.~C.,  et~al., 2021, \mn@doi [\mnras] {10.1093/mnras/stab943}, \href
  {https://ui.adsabs.harvard.edu/abs/2021MNRAS.504.1545D} {504, 1545}

\bibitem[\protect\citeauthoryear{Dolphin}{Dolphin}{2009}]{Dolphin_2009}
Dolphin A.~E.,  2009, \mn@doi [Publications of the Astronomical Society of the
  Pacific] {10.1086/600028}, 121, 655

\bibitem[\protect\citeauthoryear{{Fabbiano}}{{Fabbiano}}{1989}]{fabbiano89}
{Fabbiano} G.,  1989, \mn@doi [\araa] {10.1146/annurev.aa.27.090189.000511},
  \href {https://ui.adsabs.harvard.edu/abs/1989ARA&A..27...87F} {27, 87}

\bibitem[\protect\citeauthoryear{{Gavazzi}, {Boselli}, {Scodeggio}, {Pierini}
  \& {Belsole}}{{Gavazzi} et~al.}{1999}]{distance}
{Gavazzi} G.,  {Boselli} A.,  {Scodeggio} M.,  {Pierini} D.,   {Belsole} E.,
  1999, \mn@doi [\mnras] {10.1046/j.1365-8711.1999.02350.x}, \href
  {https://ui.adsabs.harvard.edu/abs/1999MNRAS.304..595G} {304, 595}

\bibitem[\protect\citeauthoryear{{Giersz}, {Askar}, {Wang}, {Hypki}, {Leveque}
  \& {Spurzem}}{{Giersz} et~al.}{2019}]{giersz19}
{Giersz} M.,  {Askar} A.,  {Wang} L.,  {Hypki} A.,  {Leveque} A.,   {Spurzem}
  R.,  2019, \mn@doi [\mnras] {10.1093/mnras/stz1460}, \href
  {https://ui.adsabs.harvard.edu/abs/2019MNRAS.487.2412G} {487, 2412}

\bibitem[\protect\citeauthoryear{{Giesers} et~al.,}{{Giesers}
  et~al.}{2018}]{giesers18}
{Giesers} B.,  et~al., 2018, \mn@doi [\mnras] {10.1093/mnrasl/slx203}, \href
  {https://ui.adsabs.harvard.edu/abs/2018MNRAS.475L..15G} {475, L15}

\bibitem[\protect\citeauthoryear{{Giesers} et~al.,}{{Giesers}
  et~al.}{2019}]{giesers19}
{Giesers} B.,  et~al., 2019, \mn@doi [\aap] {10.1051/0004-6361/201936203},
  \href {https://ui.adsabs.harvard.edu/abs/2019A&A...632A...3G} {632, A3}

\bibitem[\protect\citeauthoryear{{Gilfanov}}{{Gilfanov}}{2004}]{add}
{Gilfanov} M.,  2004, \mn@doi [\mnras] {10.1111/j.1365-2966.2004.07473.x},
  \href {https://ui.adsabs.harvard.edu/abs/2004MNRAS.349..146G} {349, 146}

\bibitem[\protect\citeauthoryear{{Gilfanov}, {Fabbiano}, {Lehmer}  \&
  {Zezas}}{{Gilfanov} et~al.}{2022}]{2022hxga.book..105G}
{Gilfanov} M.,  {Fabbiano} G.,  {Lehmer} B.,   {Zezas} A.,  2022, in , Handbook
  of X-ray and Gamma-ray Astrophysics.
p.~105, \mn@doi{10.1007/978-981-16-4544-0_108-1}

\bibitem[\protect\citeauthoryear{{Giordano}, {Cortese}, {Trinchieri}, {Wolter},
  {Colpi}, {Gavazzi}  \& {Mayer}}{{Giordano}
  et~al.}{2005}]{2005ApJ...634..272G}
{Giordano} L.,  {Cortese} L.,  {Trinchieri} G.,  {Wolter} A.,  {Colpi} M.,
  {Gavazzi} G.,   {Mayer} L.,  2005, \mn@doi [\apj] {10.1086/496942}, \href
  {https://ui.adsabs.harvard.edu/abs/2005ApJ...634..272G} {634, 272}

\bibitem[\protect\citeauthoryear{{Humphrey} \& {Buote}}{{Humphrey} \&
  {Buote}}{2008}]{2008ApJ...689..983H}
{Humphrey} P.~J.,  {Buote} D.~A.,  2008, \mn@doi [\apj] {10.1086/592590}, \href
  {https://ui.adsabs.harvard.edu/abs/2008ApJ...689..983H} {689, 983}

\bibitem[\protect\citeauthoryear{{Hunt}, {Gallo}, {Chandar}, {Mok}  \&
  {Prestwich}}{{Hunt} et~al.}{2022}]{2022arXiv220607192H}
{Hunt} Q.,  {Gallo} E.,  {Chandar} R.,  {Mok} A.,   {Prestwich} A.,  2022,
  \mn@doi [arXiv e-prints] {10.48550/arXiv.2206.07192}, \href
  {https://ui.adsabs.harvard.edu/abs/2022arXiv220607192H} {p. arXiv:2206.07192}

\bibitem[\protect\citeauthoryear{{Irwin}}{{Irwin}}{2005}]{2005ApJ...631..511I}
{Irwin} J.~A.,  2005, \mn@doi [\apj] {10.1086/432611}, \href
  {https://ui.adsabs.harvard.edu/abs/2005ApJ...631..511I} {631, 511}

\bibitem[\protect\citeauthoryear{{Irwin}, {Brink}, {Bregman}  \&
  {Roberts}}{{Irwin} et~al.}{2010}]{Irwin}
{Irwin} J.~A.,  {Brink} T.~G.,  {Bregman} J.~N.,   {Roberts} T.~P.,  2010,
  \mn@doi [\apjl] {10.1088/2041-8205/712/1/L1}, \href
  {https://ui.adsabs.harvard.edu/abs/2010ApJ...712L...1I} {712, L1}

\bibitem[\protect\citeauthoryear{{Jester} et~al.,}{{Jester}
  et~al.}{2005}]{Jester05}
{Jester} S.,  et~al., 2005, \mn@doi [\aj] {10.1086/432466}, \href
  {https://ui.adsabs.harvard.edu/abs/2005AJ....130..873J} {130, 873}

\bibitem[\protect\citeauthoryear{{Kim}, {Kim}, {Fabbiano}, {Lee}, {Park},
  {Geisler}  \& {Dirsch}}{{Kim} et~al.}{2006}]{2006ApJ...647..276K}
{Kim} E.,  {Kim} D.-W.,  {Fabbiano} G.,  {Lee} M.~G.,  {Park} H.~S.,  {Geisler}
  D.,   {Dirsch} B.,  2006, \mn@doi [\apj] {10.1086/505261}, \href
  {https://ui.adsabs.harvard.edu/abs/2006ApJ...647..276K} {647, 276}

\bibitem[\protect\citeauthoryear{{Kim}, {Fabbiano}, {Ivanova}, {Fragos},
  {Jord{\'a}n}, {Sivakoff}  \& {Voss}}{{Kim}
  et~al.}{2013}]{2013ApJ...764...98K}
{Kim} D.~W.,  {Fabbiano} G.,  {Ivanova} N.,  {Fragos} T.,  {Jord{\'a}n} A.,
  {Sivakoff} G.~R.,   {Voss} R.,  2013, \mn@doi [\apj]
  {10.1088/0004-637X/764/1/98}, \href
  {https://ui.adsabs.harvard.edu/abs/2013ApJ...764...98K} {764, 98}

\bibitem[\protect\citeauthoryear{{King}, {Lasota}  \& {Middleton}}{{King}
  et~al.}{2023}]{2023NewAR..9601672K}
{King} A.,  {Lasota} J.-P.,   {Middleton} M.,  2023, \mn@doi [\nar]
  {10.1016/j.newar.2022.101672}, \href
  {https://ui.adsabs.harvard.edu/abs/2023NewAR..9601672K} {96, 101672}

\bibitem[\protect\citeauthoryear{{Kirsten} et~al.,}{{Kirsten}
  et~al.}{2022}]{2022Natur.602..585K}
{Kirsten} F.,  et~al., 2022, \mn@doi [\nat] {10.1038/s41586-021-04354-w}, \href
  {https://ui.adsabs.harvard.edu/abs/2022Natur.602..585K} {602, 585}

\bibitem[\protect\citeauthoryear{{Kovlakas}, {Zezas}, {Andrews}, {Basu-Zych},
  {Fragos}, {Hornschemeier}, {Lehmer}  \& {Ptak}}{{Kovlakas}
  et~al.}{2020}]{kovlakas}
{Kovlakas} K.,  {Zezas} A.,  {Andrews} J.~J.,  {Basu-Zych} A.,  {Fragos} T.,
  {Hornschemeier} A.,  {Lehmer} B.,   {Ptak} A.,  2020, \mn@doi [\mnras]
  {10.1093/mnras/staa2481}, \href
  {https://ui.adsabs.harvard.edu/abs/2020MNRAS.498.4790K} {498, 4790}

\bibitem[\protect\citeauthoryear{{Kremer}, {Chatterjee}, {Ye}, {Rodriguez}  \&
  {Rasio}}{{Kremer} et~al.}{2019}]{kremer19}
{Kremer} K.,  {Chatterjee} S.,  {Ye} C.~S.,  {Rodriguez} C.~L.,   {Rasio}
  F.~A.,  2019, \mn@doi [\apj] {10.3847/1538-4357/aaf646}, \href
  {https://ui.adsabs.harvard.edu/abs/2019ApJ...871...38K} {871, 38}

\bibitem[\protect\citeauthoryear{{Kundu}, {Maccarone}, {Zepf}  \&
  {Puzia}}{{Kundu} et~al.}{2003}]{Kundu2003}
{Kundu} A.,  {Maccarone} T.~J.,  {Zepf} S.~E.,   {Puzia} T.~H.,  2003, \mn@doi
  [\apjl] {10.1086/376493}, \href
  {https://ui.adsabs.harvard.edu/abs/2003ApJ...589L..81K} {589, L81}

\bibitem[\protect\citeauthoryear{{Kundu}, {Maccarone}  \& {Zepf}}{{Kundu}
  et~al.}{2007}]{2007ApJ...662..525K}
{Kundu} A.,  {Maccarone} T.~J.,   {Zepf} S.~E.,  2007, \mn@doi [\apj]
  {10.1086/518021}, \href
  {https://ui.adsabs.harvard.edu/abs/2007ApJ...662..525K} {662, 525}

\bibitem[\protect\citeauthoryear{{Lehmer}, {Alexander}, {Bauer}, {Brandt},
  {Goulding}, {Jenkins}, {Ptak}  \& {Roberts}}{{Lehmer}
  et~al.}{2010}]{2010ApJ...724..559L}
{Lehmer} B.~D.,  {Alexander} D.~M.,  {Bauer} F.~E.,  {Brandt} W.~N.,
  {Goulding} A.~D.,  {Jenkins} L.~P.,  {Ptak} A.,   {Roberts} T.~P.,  2010,
  \mn@doi [\apj] {10.1088/0004-637X/724/1/559}, \href
  {http://adsabs.harvard.edu/abs/2010ApJ...724..559L} {724, 559}

\bibitem[\protect\citeauthoryear{{Lehmer} et~al.,}{{Lehmer}
  et~al.}{2020}]{Lehmer20}
{Lehmer} B.~D.,  et~al., 2020, \mn@doi [\apjs] {10.3847/1538-4365/ab9175},
  \href {https://ui.adsabs.harvard.edu/abs/2020ApJS..248...31L} {248, 31}

\bibitem[\protect\citeauthoryear{{Lemons}, {Reines}, {Plotkin}, {Gallo}  \&
  {Greene}}{{Lemons} et~al.}{2015}]{Lemons15}
{Lemons} S.~M.,  {Reines} A.~E.,  {Plotkin} R.~M.,  {Gallo} E.,   {Greene}
  J.~E.,  2015, \mn@doi [\apj] {10.1088/0004-637X/805/1/12}, \href
  {https://ui.adsabs.harvard.edu/abs/2015ApJ...805...12L} {805, 12}

\bibitem[\protect\citeauthoryear{{Leveque}, {Giersz}, {Askar}  \&
  {Arca-Sedda}}{{Leveque} et~al.}{2022}]{2022arXiv220901564L}
{Leveque} A.,  {Giersz} M.,  {Askar} A.,   {Arca-Sedda} M.,  2022, arXiv
  e-prints, \href {https://ui.adsabs.harvard.edu/abs/2022arXiv220901564L} {p.
  arXiv:2209.01564}

\bibitem[\protect\citeauthoryear{{Maccarone} \& {Knigge}}{{Maccarone} \&
  {Knigge}}{2007}]{mk07}
{Maccarone} T.,  {Knigge} C.,  2007, \mn@doi [Astronomy and Geophysics]
  {10.1111/j.1468-4004.2007.48512.x}, \href
  {https://ui.adsabs.harvard.edu/abs/2007A&G....48e..12M} {48, 5.12}

\bibitem[\protect\citeauthoryear{{Maccarone}, {Kundu}, {Zepf}  \&
  {Rhode}}{{Maccarone} et~al.}{2007}]{2007Natur.445..183M}
{Maccarone} T.~J.,  {Kundu} A.,  {Zepf} S.~E.,   {Rhode} K.~L.,  2007, \mn@doi
  [\nat] {10.1038/nature05434}, \href
  {https://ui.adsabs.harvard.edu/abs/2007Natur.445..183M} {445, 183}

\bibitem[\protect\citeauthoryear{{Maccarone}, {Kundu}, {Zepf}  \&
  {Rhode}}{{Maccarone} et~al.}{2011}]{Maccarone11}
{Maccarone} T.~J.,  {Kundu} A.,  {Zepf} S.~E.,   {Rhode} K.~L.,  2011, \mn@doi
  [\mnras] {10.1111/j.1365-2966.2010.17547.x}, \href
  {https://ui.adsabs.harvard.edu/abs/2011MNRAS.410.1655M} {410, 1655}

\bibitem[\protect\citeauthoryear{{Miller-Jones} et~al.,}{{Miller-Jones}
  et~al.}{2015}]{MillerJones15}
{Miller-Jones} J.~C.~A.,  et~al., 2015, \mn@doi [\mnras]
  {10.1093/mnras/stv1869}, \href
  {http://adsabs.harvard.edu/abs/2015MNRAS.453.3918M} {453, 3918}

\bibitem[\protect\citeauthoryear{{Morscher}, {Pattabiraman}, {Rodriguez},
  {Rasio}  \& {Umbreit}}{{Morscher} et~al.}{2015}]{Morscher}
{Morscher} M.,  {Pattabiraman} B.,  {Rodriguez} C.,  {Rasio} F.~A.,   {Umbreit}
  S.,  2015, \mn@doi [\apj] {10.1088/0004-637X/800/1/9}, \href
  {https://ui.adsabs.harvard.edu/abs/2015ApJ...800....9M} {800, 9}

\bibitem[\protect\citeauthoryear{{Peacock}, {Maccarone}, {Kundu}  \&
  {Zepf}}{{Peacock} et~al.}{2010}]{Peacock2010}
{Peacock} M.~B.,  {Maccarone} T.~J.,  {Kundu} A.,   {Zepf} S.~E.,  2010,
  \mn@doi [\mnras] {10.1111/j.1365-2966.2010.17119.x}, \href
  {https://ui.adsabs.harvard.edu/abs/2010MNRAS.407.2611P} {407, 2611}

\bibitem[\protect\citeauthoryear{{Peacock}, {Zepf}, {Kundu}, {Maccarone},
  {Lehmer}, {Gonzalez}  \& {Maraston}}{{Peacock}
  et~al.}{2017a}]{2017MNRAS.466.4021P}
{Peacock} M.~B.,  {Zepf} S.~E.,  {Kundu} A.,  {Maccarone} T.~J.,  {Lehmer}
  B.~D.,  {Gonzalez} A.~H.,   {Maraston} C.,  2017a, \mn@doi [\mnras]
  {10.1093/mnras/stw3375}, \href
  {https://ui.adsabs.harvard.edu/abs/2017MNRAS.466.4021P} {466, 4021}

\bibitem[\protect\citeauthoryear{{Peacock} et~al.,}{{Peacock}
  et~al.}{2017b}]{Peacock17}
{Peacock} M.~B.,  et~al., 2017b, \mn@doi [\apj] {10.3847/1538-4357/aa70eb},
  \href {https://ui.adsabs.harvard.edu/abs/2017ApJ...841...28P} {841, 28}

\bibitem[\protect\citeauthoryear{{Roberts} et~al.,}{{Roberts}
  et~al.}{2012}]{2012ApJ...760..135R}
{Roberts} T.~P.,  et~al., 2012, \mn@doi [\apj] {10.1088/0004-637X/760/2/135},
  \href {https://ui.adsabs.harvard.edu/abs/2012ApJ...760..135R} {760, 135}

\bibitem[\protect\citeauthoryear{{Rodriguez}, {Chatterjee}  \&
  {Rasio}}{{Rodriguez} et~al.}{2016}]{rodriguez2016}
{Rodriguez} C.~L.,  {Chatterjee} S.,   {Rasio} F.~A.,  2016, \mn@doi [\prd]
  {10.1103/PhysRevD.93.084029}, \href
  {http://adsabs.harvard.edu/abs/2016PhRvD..93h4029R} {93, 084029}

\bibitem[\protect\citeauthoryear{{Sarazin}, {Irwin}  \& {Bregman}}{{Sarazin}
  et~al.}{2001}]{2001ApJ...556..533S}
{Sarazin} C.~L.,  {Irwin} J.~A.,   {Bregman} J.~N.,  2001, \mn@doi [\apj]
  {10.1086/321618}, \href
  {https://ui.adsabs.harvard.edu/abs/2001ApJ...556..533S} {556, 533}

\bibitem[\protect\citeauthoryear{{Sathyaprakash} et~al.,}{{Sathyaprakash}
  et~al.}{2022}]{2022MNRAS.511.5346S}
{Sathyaprakash} R.,  et~al., 2022, \mn@doi [\mnras] {10.1093/mnras/stac402},
  \href {https://ui.adsabs.harvard.edu/abs/2022MNRAS.511.5346S} {511, 5346}

\bibitem[\protect\citeauthoryear{{Shih}, {Kundu}, {Maccarone}, {Zepf}  \&
  {Joseph}}{{Shih} et~al.}{2010}]{Shih}
{Shih} I.~C.,  {Kundu} A.,  {Maccarone} T.~J.,  {Zepf} S.~E.,   {Joseph} T.~D.,
   2010, \mn@doi [\apj] {10.1088/0004-637X/721/1/323}, \href
  {https://ui.adsabs.harvard.edu/abs/2010ApJ...721..323S} {721, 323}

\bibitem[\protect\citeauthoryear{{Sivakoff} et~al.,}{{Sivakoff}
  et~al.}{2007}]{2007ApJ...660.1246S}
{Sivakoff} G.~R.,  et~al., 2007, \mn@doi [\apj] {10.1086/513094}, \href
  {https://ui.adsabs.harvard.edu/abs/2007ApJ...660.1246S} {660, 1246}

\bibitem[\protect\citeauthoryear{{Spitzer}}{{Spitzer}}{1969}]{1969ApJ...158L.139S}
{Spitzer} Lyman J.,  1969, \mn@doi [\apjl] {10.1086/180451}, \href
  {https://ui.adsabs.harvard.edu/abs/1969ApJ...158L.139S} {158, L139}

\bibitem[\protect\citeauthoryear{{Sridhar}, {Metzger}, {Beniamini}, {Margalit},
  {Renzo}, {Sironi}  \& {Kovlakas}}{{Sridhar}
  et~al.}{2021}]{2021ApJ...917...13S}
{Sridhar} N.,  {Metzger} B.~D.,  {Beniamini} P.,  {Margalit} B.,  {Renzo} M.,
  {Sironi} L.,   {Kovlakas} K.,  2021, \mn@doi [\apj]
  {10.3847/1538-4357/ac0140}, \href
  {https://ui.adsabs.harvard.edu/abs/2021ApJ...917...13S} {917, 13}

\bibitem[\protect\citeauthoryear{{Strader}, {Chomiuk}, {Maccarone},
  {Miller-Jones}  \& {Seth}}{{Strader} et~al.}{2012}]{strader12}
{Strader} J.,  {Chomiuk} L.,  {Maccarone} T.~J.,  {Miller-Jones} J. C.~A.,
  {Seth} A.~C.,  2012, \mn@doi [\nat] {10.1038/nature11490}, \href
  {https://ui.adsabs.harvard.edu/abs/2012Natur.490...71S} {490, 71}

\bibitem[\protect\citeauthoryear{{Swartz}, {Tennant}  \& {Soria}}{{Swartz}
  et~al.}{2009}]{swartz}
{Swartz} D.~A.,  {Tennant} A.~F.,   {Soria} R.,  2009, \mn@doi [\apj]
  {10.1088/0004-637X/703/1/159}, \href
  {https://ui.adsabs.harvard.edu/abs/2009ApJ...703..159S} {703, 159}

\bibitem[\protect\citeauthoryear{{Weatherford}, {Chatterjee}, {Kremer}  \&
  {Rasio}}{{Weatherford} et~al.}{2020}]{2020ApJ...898..162W}
{Weatherford} N.~C.,  {Chatterjee} S.,  {Kremer} K.,   {Rasio} F.~A.,  2020,
  \mn@doi [\apj] {10.3847/1538-4357/ab9f98}, \href
  {https://ui.adsabs.harvard.edu/abs/2020ApJ...898..162W} {898, 162}

\makeatother
\end{thebibliography}
%\begin{thebibliography}{99}
%\end{thebibliography}

%%%%%%%%%%%%%%%%%%%%%%%%%%%%%%%%%%%%%%%%%%%%%%%%%%

%%%%%%%%%%%%%%%%% APPENDICES %%%%%%%%%%%%%%%%%%%%%

\appendix
\section{Additional Tables}
We list below the additional detected X-ray sources in ObsID 834 and ObsID 9569 that were not matched to a globular cluster from the catalogue of \citet{Bonfini}.
\begin{table*}
	\centering
	\caption{X-ray and optical properties of non-GC X-ray point sources identified in NGC 4261 with their positions indicated (GC X-ray sources are listed in Tables \ref{tab:ulx} and \ref{tab:lmxb}). Sources that fall in the field of view for the HST catalog are marked with 'A' and those that have been identified in \citealt{2005ApJ...634..272G} are denoted with 'B' .  There are 98 sources including the GCLMXB and ULX population.  X-ray luminosities are in the 0.5-8.0 keV band. The variability column indicates if the source varies beyond the measurement uncertainties if measured in both observations. }
	\label{tab:xray1}
	\begin{tabular}{lcccccccr} % four columns, alignment for each
		\hline
		\hline
		Name & RA & Dec & Counts (834) & Counts(9569) & L$_X$  (834) & L$_X$  (9569)  &  HST Field/Giordano & Variability \\
		&&& 0.5-8.0 keV & 0.5-8.0 keV & $10^{38}$ erg s$^{-1}$ & $10^{38}$ erg s$^{-1}$ & & (Y/N) \\
		\hline
GCULX1 &  12:19:24.631 &  +05:51:04.86 & $28.41 \pm 5.56$ & $71.72 \pm 8.60$  & $2.0^{+0.7}_{-0.6}$ &$2.3 \pm 0.4$ &A/B & N \\
&&&&&&&&\\
GCULX2 & 12:19:20.260 & +05:49:08.43 & $23.55 \pm 5.09$  & $23.19\pm5.00$ & $0.4^{+0.2}_{-0.1}$&$0.8 \pm 0.2$ & A/B & Y \\
&&&&&&&&\\
GCLMXB1 & 12:19:25.512 & +05:47:41.86 & $18.61 \pm 4.47$& $6.89\pm2.82$ & $3.2^{+1.4}_{-1.1}$ & $3.0^{+1.3}_{-1.0}$  & A/B & N\\
&&&&&&&&\\
GCLMXB2 & 12:19:23.598 & +05:47:55.48 & $13.03\pm 3.87$ &-&$5.3^{+2.9}_{-2.2}$ & - &A/B & N\\
&&&&&&&&\\
GCLMXB3 & 12:19:21.092 & +05:48:07.09  &$9.56\pm3.46$ & - & $2.0^{+1.5}_{-1.1}$ &-   & A/B & N\\
&&&&&&&&\\
GCLMXB4 & 12:19:20.582 & +05:48:35.67 & $15.35\pm4.12$  &- & $4.9^{+2.5}_{-1.9}$ & -& A/B & N\\
&&&&&&&&\\
GCLMXB5& 12:19:22.096 & +05:48:40.83 &$5.78\pm2.64$ & - & $1.4^{+1.2}_{-0.8}$ & - & A/B & N\\ %double match
&&&&&&&&\\
GCLMXB6 & 12:19:21.003 & +05:48:44.33 &$12.85\pm3.74$  & $10.76 \pm 3.60$ & $2.8^{+1.5}_{-1.1}$ & $3.1^{+1.0}_{-0.9}$&A/B & N\\
&&&&&&&&\\
GCLMXB7 & 12:19:22.367 & +05:48:47.98 & $11.29\pm3.60$ &- & $3.5^{+2.0}_{-1.5}$ &- & A/B & N\\
&&&&&&&&\\
GCLMXB8 & 12:19:20.956  & +05:49:26.81 & $16.52\pm4.35$  & $11.40\pm3.60$& $3.4^{+1.6}_{-1.2}$ & $3.2^{+1.0}_{-0.8}$   &A/B & N\\
&&&&&&&&\\
GCLMXB9 & 12:19:20.670 & +05:49:29.20  & $7.57\pm 3.00$ & $13.06\pm3.87$ & $1.5^{+1.2}_{-0.8}$ &$3.5^{+1.0}_{-0.9}$ & A/B & Y\\
&&&&&&&&\\
GCLMXB10& 12:19:24.936 & +05:49:32.23 & $12.91\pm4.00$  & $8.27\pm3.16$ & $3.1^{+1.7}_{-1.3}$ & $4.4^{+1.3}_{-1.1}$  &A/B & N\\
&&&&&&&&\\
GCLMXB11& 12:19:24.444 & +05:49:33.04 & $7.16\pm3.31$ & - &$2.0^{+1.3}_{-0.9}$&- & A/B & N\\ 
&&&&&&&&\\
GCLMXB12 & 12:19:18.068 & +05:49:54.10 & $9.89\pm3.31$ &$7.85\pm3.00$ & $1.4^{+1.2}_{-0.8}$& $2.7^{+0.8}_{-1.1}$ & A/B & N \\
&&&&&&&&\\
GCLMXB13& 12:19:24.842 & +05:50:08.59 & $9.21\pm3.46$ & $14.85\pm4.24$& $3.9^{+2.9}_{-2.2}$ &$3.7^{+1.2}_{-1.1}$  & A/B & N\\
&&&&&&&&\\
GCLMXB14 & 12:19:23.236 & +05:50:13.49& $9.98\pm3.60$  &$5.91\pm2.64$ &$1.6^{+1.2}_{-0.9}$ &$1.9^{+1.0}_{-1.1}$   & A/B & N\\
&&&&&&&&\\
GCLMXB15 & 12:19:26.914 & +05:50:44.21 & $18.67\pm4.47$  & - & $4.7^{+2.7}_{-1.6}$&-&A/B & N\\
&&&&&&&&\\
GCLMXB16 & 12:19:22.167 &+05:50:59.41  & $8.01\pm3.00$ &$7.01\pm2.82$ & $2.3^{+1.6}_{-1.1}$ &$2.1^{+0.7}_{-1.1}$ & A/B & N\\
&&&&&&&&\\
GCLMXB17 & 12:19:24.759 & +05:50:14.50 & - & $6.03\pm2.64$& - & $2.6^{+1.6}_{-1.2}$ &A/B & N\\
&&&&&&&&\\
GCLMXB18 & 12:19:24.978 &+05:50:58.28 & - &$7.00\pm3.16$ &- &$3.6^{+1.2}_{-1.0}$ &A/B & N \\
&&&&&&&&\\
1  &  12:19:21.217& +5:47:52.11 & $17.05 \pm 4.47$ & -  &   $3.8^{+1.5}_{ -1.9}$ &    - &       A/B  & -\\
		&&&&&&&&\\
2  &  12:19:24.252 &   +5:47:54.72 & $16.92  \pm 4.36$ &    - &   $3.2^{+1.1}_{-1.5}$ &    - &      A/B & - \\
&&&&&&&&\\
3  &  12:19:23.59 &  +5:47:55.64 & $13.04 \pm 3.87$ & - &   $5.3^{+2.2}_{-2.9}$ &    - &      A/B & -\\
&&&&&&&&\\
4  &  12:19:23.18 &   +5:48:07.12 &    $ 9.13 \pm 3.16$ &     - &    $1.6^{+0.7}_{-  1.0}$ &   -  &   A/B & - \\
&&&&&&&&\\
5  &  12:19:29.95 &  +5:50:27.95 &     $20.91 \pm       4.79$ &     $13.39 \pm     4.12$  &   $5.7^{+1.8}_{-2.3}$ &    $5.4^{+1.5}_{-1.7}$ &  B & N\\
&&&&&&&&\\
6  &  12:19:21.06 &  +5:48:07.14 & $9.56 \pm 3.46 $ &-&   $2.0^{+1.1}_{-1.5}$ &    - &       A/B  & -\\
&&&&&&&&\\
7  &  12:19:19.77 &  +5:48:22.30 &$11.21 \pm 3.60$ &- &    $1.5^{+0.7}_{-0.9}$ &   - &      A/B & - \\
&&&&&&&&\\
8  &  12:19:21.55 &   +5:50:36.31 &  - & $10.00 \pm 3.31$ &    $3.4^{+1.7}_{-2.3}$ &-  &     A &-\\
&&&&&&&&\\
9 &  12:19:20.60 &   +5:48:35.92 &$15.35 \pm 4.12$ &  - &    $4.9^{+1.9}_{_2.5}$ &  - &      A/B &  -\\
&&&&&&&&\\
10  &  12:19:35.90&   +5:50:44.87 & -  &    $ 11.42 \pm 3.74 $ &  - &   $ 3.4^{+1.2}_{  -1.5}$ &  - & -\\

\hline
\end{tabular}
\end{table*}

\begin{table*}
	\centering
	\label{tab:xra-2}
	\begin{tabular}{lcccccccr} % four columns, alignment for each
		\hline
		\hline
		Name & RA & Dec & Counts (834) & Counts(9569) & L$_X$  (834) & L$_X$  (9569)  &  HST Field/Giordano & Variability \\
		&&& 0.5-8.0 keV & 0.5-8.0 keV & $10^{38}$ erg s$^{-1}$ & $10^{38}$ erg s$^{-1}$ & & (Y/N) \\
		\hline

11 &  12:19:22.14  &  +5:48:41.00 & $5.78 \pm 2.64$ &   - &   $1.4^{+0.8}_ {-1.2}$ &  -  &       A/B &  -\\
&&&&&&&&\\
12 &  12:19:35.63 &   +5:50:46.68 &    -& $8.30 \pm 3.31$ &  - & $1.8^{+0.9}_{-1.2}$ &  - & -\\
&&&&&&&&\\
13 &  12:19:22.36 &  +5:48:48.03 & $11.29 \pm 3.60$ & - &   $3.5^{+1.5}_{-2.0}$ &   -  &      A/B &-\\
&&&&&&&&\\
14 &  12:19:20.43 &  +5:48:02.79 &   $13.45 \pm 3.87$ &     $ 6.65 \pm 2.83$ &    $4.1^{+1.6}_{-2.1}$ &    $2.9^{+1.0}_{-1.3}$ &   A/B & N\\
&&&&&&&&\\
15 &  12:19:19.18 &   +5:48:49.24 & $7.86 \pm 3.16$ & - &    $2.9^{+1.5}_{-2.1}$ &   - &      A/B & -\\
&&&&&&&&\\
16 &  12:19:19.92 &   +5:48:51.05 &$13.28  \pm 3.74$ &    - &    $2.3^{+0.9}_{-1.2}$ &  - &     A & - \\
&&&&&&&&\\
17 &  12:19:14.84 &   +5:51:17.52 &-  &     $49.57 \pm 7.28$ &  - &   $16.9^{+2.4}_{- 2.4}$ &      A &- \\
&&&&&&&&\\
18 &  12:19:19.32 &   +5:49:06.41 & $ 7.10 \pm 2.82$ &   - &   $3.8^{+2.0}_{-3.0}$ &   - &     A/B & -\\
&&&&&&&&\\
19 &  12:19:17.48 &  +5:51:57.03 & - &     $37.65 \pm 6.32$ & - &    $9.2^{2.1}_{-2.1}$ & A &-\\
&&&&&&&&\\
20 &  12:19:15.50 &   +5:52:33.07  & - &     $15.35 \pm 4.00$ &   - &   $11.8^{+2.2}_{-2.2}$ & - &- \\
&&&&&&&&\\
21 &  12:19:13.09 &  +5:49:9.19 & 5.31 $\pm$  2.45 &     -&  $6.9^{+4.1}_{-6.4}$ &    - &       - &- \\
&&&&&&&&\\
22 &  12:19:21.17 &  +5:44:22.61 &- &    $ 24.17  \pm 6.08$ &    -&   $10.1^{+3.0}_{-3.4}$ &      - &-\\
&&&&&&&&\\
23 &  12:19:10.26 &   +5:44:52.39 &-&  $32.99 \pm 7.55$ & - &   $21.8^{+3.0}_{-3.0}$ &  - & -\\
&&&&&&&&\\
24 &  12:19:09.14 &  +5:44:59.03 &  - & $48.23 \pm 8.43$ &  - &   $27.1 ^{+3.0}_{-3.0}$ & - &-\\
&&&&&&&&\\
25 &  12:19:27.19 &   +5:46:12.88 & - &     $11.27         \pm 3.74$ &    -&    $2.7   ^{+0.9}_{-1.1}$ &   - & - \\
&&&&&&&&\\
26 &  12:19:22.69 &   +5:49:43.47 &$15.981 \pm 5.47$ &    $ 16.63  \pm  4.35$ &    $3.8^{+3.0}_{-3.6}$ &    $5.0^{+1.5}_{-1.7}$ &   A &N\\
&&&&&&&&\\
27 &  12:19:22.72 &   +5:47:22.35 &-& $8.39 \pm         3.16$ &   - &    $1.9   ^{+0.7}_{-0.8}$ &       A/B & -\\
&&&&&&&&\\
28 &  12:19:22.81 &   +5:47:35.89 & $13.11 \pm 3.87$ &     $10.06 \pm 3.46$ &   $3.5^    {+1.6}_{-2.1}$ &    $4.5^{+1.4}_{-1.7}$ &  A & N \\
&&&&&&&&\\
29 &  12:19:12.39 &   +5:48:47.67 &-&     $10.19 \pm 3.60$ &  - &$3.0^{+1.6}  _{-2.0}$ &  - & -\\
&&&&&&&&\\
30 &  12:19:24.43 &   +5:49:33.53 &  $7.16  \pm 3.31$&  -&    $2.0^{+0.9}_{-1.3}$ &    - &     A & -\\
	&&&&&&&&\\
31 &  12:19:23.49 &   +5:49:36.09 & $32.10 \pm 9.11$ &  -&   $8.2^{+4.2}_{ -4.6}$ &-& A/B & -\\
&&&&&&&&\\
32 &  12:19:16.78 &   
+5:50:18.02 & - &      $7.09 \pm       3.00$ &   - &   $2.6^{+1.6}_{ -2.3}$ & - & -\\
&&&&&&&&\\
33 &  12:19:21.86 &   5:50:09.03 &    $7.23 \pm 3.00$ &     $16.41 \pm 4.36$ & $2.7^{+1.6} _{-2.2}$ &    $4.91^{+1.32}_{-1.58}$ &   A & Y \\
&&&&&&&&\\
34 &  12:19:38.55 &   +5:50:53.65 &     - &$5.67 \pm 2.64$ & -&    $1.9^{+0.9}_{-1.1}$ &  -&- \\
&&&&&&&&\\
35 &  12:19:21.98 & +5:49:47.09   &  $7.59 \pm 3.16$ &   - &$3.3^{+1.6}_{-2.3}$ &   - & A/B &- \\
&&&&&&&&\\
36 &  12:19:32.10 &  +5:50:30.10 &    $6.59 \pm 2.83$ & $6.92 \pm 2.82$  & $4.1^{+2.2}_{ -3.2}$ &    $5.4^{+1.8}_{-2.3}$ &   - & N\\
&&&&&&&&\\
37 &  12:19:19.75 &   +5:51:48.54 &    - &  $10.46\pm 3.60$ &  - & $2.4^{+1.5}_{-1.9}$ &  - & -\\
&&&&&&&&\\
38 &  12:19:29.21 &   +5:50:0.812 &  $10.13 \pm 3.46$ &    - &$4.1^{+1.9}_{  -2.5}$ &    - & B & -\\
&&&&&&&&\\
39 &  12:19:24.05 & +5:50:26.99 & $22.50\pm 6.40$ & $14.20 \pm     4.36$ & $8.8^{+4.2}_{-4.8}$ & $3.7^{+1.5}_ {-1.6}$ &  A & N\\
&&&&&&&&\\
40 &  12:19:30.51 & +5:50:7.88   &$10.34  \pm 3.31$ &      - & $5.5^{+2.3}_{-3.2}$ &    - &  B &-\\
&&&&&&&&\\
41 &  12:19:24.83 &   +5:50:8.99 & $9.21\pm 3.46$ &      - &$3.9^{+2.2}_{-3.0}$ &    - &  A/ B & -\\
&&&&&&&&\\
42 &  12:19:25.67 &   +5:50:28.69 &$10.64 \pm   3.46$ &     -&  $2.0^{+0.9}_{-1.2}$ &    - &      A/B &- \\

\hline
\end{tabular}
\end{table*}

\begin{table*}
	\centering
	\label{tab:xray2}
	\begin{tabular}{lcccccccr} % four columns, alignment for each
		\hline
		\hline
		Name & RA & Dec & Counts (834) & Counts(9569) & L$_X$  (834) & L$_X$  (9569)  &  HST Field/Giordano & Variability \\
		&&& 0.5-8.0 keV & 0.5-8.0 keV & $10^{38}$ erg s$^{-1}$ & $10^{38}$ erg s$^{-1}$ & & (Y/N) \\
		\hline

43 &  12:19:33.93 &   +5:50:35.70 &$19.28  \pm 4.69$ &- & $3.3^{+1.1}_{-1.4}$ &    - &   B &- \\
&&&&&&&&\\
44 &  12:19:26.89 &   +5:50:44.31 &$18.67\pm   4.47$ &  - & $4.7^{+1.6}_{-2.1}$ &    - &  A/ B &- \\
&&&&&&&&\\
45 &  12:19:16.81 &   +5:49:50.53 &    $34.16 \pm 6.00$ & $11.58  \pm 3.60$ &$6.8^{+1.8}_{-   2.2}$ & $5.7^{+1.2}_{-1.3}$ &   A/B &N\\
&&&&&&&&\\
46 &  12:19:31.97 &   +5:51:54.11 &$16.39 \pm 4.58$ & -& $4.1^{+1.7}_{-2.2}$ &    - &      B &- \\
&&&&&&&&\\
47 &  12:19:25.96 &  +5:52:20.50 & $8.64  \pm 3.162$ &  - & $2.8^{+1.4}_{-1.9}$ &   - & B &- \\
&&&&&&&&\\
48 &  12:19:27.03 &   +5:47:20.87 & $10.82 \pm 3.60$ &  - &  $4.8^{+2.5}_{-3.4}$ &    - &A &- \\
&&&&&&&&\\
49 &  12:19:22.11 &   +5:49:29.16 & $25.03 \pm 6.63$ & -& $5.6^{+2.1}_{-2.5}$ &    - & A &-\\
&&&&&&&&\\
50 &  12:19:21.08 &   +5:50:17.21 &  $8.39 \pm 3.16$ & -&  $5.0^{+2.6}_{- 3.6}$ &    -&A/B &- \\
&&&&&&&&\\

51&  12:19:35.52 &   +5:50:48.76 & $12.16 \pm 4.00$ &    - & $5.2^{+2.3}_{-3.0}$ &    - & B &- \\
&&&&&&&&\\
52 &  12:19:35.39 &   +5:54:02.696 &$30.74 \pm 6.71$ &  -&  $17.0^{+4.9}_{- 5.7}$ &   - &- &- \\
&&&&&&&&\\
53 &  12:19:22.81 &   +5:46:44.92 &  $12.58\pm 3.60$ &$16.50 \pm 4.36$ &$8.2^{+3.0}_{-3.9}$ & $5.1^{+1.4}_{-1.6}$ & B & N \\
&&&&&&&&\\
54 &  12:19:22.00 &   +5:50:13.89 &$9.74\pm 3.46$ & $13.91  \pm 4.00$ &$2.4^{+1.1}_{-1.6}$ &$3.2^{+1.0}_{-1.2}$ &  A &N\\
&&&&&&&&\\
55 &  12:19:26.06 &   +5:50:13.89 & $7.74   \pm 3.00$ &     $24.99 \pm 5.19$ & $1.7^{+0.9}_{-1.2}$ & $10.4^{+1.8}_ {-1.8}$& A/B& Y\\
&&&&&&&&\\
56 &  12:19:18.23 &   +5:49:12.17 & $8.01\pm 3.00$ & $10.03\pm 3.46$ & $7.6^{+3.5}_{-5.1}$&$1.9^{+0.7}_{- 0.9}$& A/B &Y\\
&&&&&&&&\\
57 &  12:19:29.78 &   +5:51:14.42 &    $23.10 \pm 5.00$ &$13.32\pm 4.00$&$5.9^{+1.8}_{-2.3}$ & $4.4^{+1.1}_{-1.3} $&   B &N \\
&&&&&&&&\\
58 &  12:19:23.12 &   +5:47:41.25 &   $19.56 \pm 4.89$ & $15.73\pm 4.24$ &  $4.6^{+1.7}_{-  2.1}$ & $5.0^{+1.3}_{-1.3}$ &   A/B & N\\
&&&&&&&&\\
59 &  12:19:23.20 & +5:49:29.86 &$2931.73 \pm 65.23$& $1834.08 \pm 43.87$ &  $785.0^{+23.6}_{-  23.7}$ & $602.8^{+16.4}_{-16.5}$ &   A/B & Y \\
&&&&&&&&\\
60 &  12:19:31.08 &  +5:50:8.15 & $11.69 \pm     3.60$ & $14.56 \pm 4.12$ &$3.6^{+1.5}_{-2.1}$ &$4.9^{+1.2}_{-1.2}$ &  B & N\\
&&&&&&&&\\
61 &  12:19:26.33 &   +5:46:16.11 &     - &      $5.72 \pm 2.64 $&  - &$4.4^{+3.0}_{-4.5}$ &  - &-\\
&&&&&&&&\\
62 &  12:19:17.62 &  +5:47:28.54 &      - &$9.45\pm 3.46$ &  - &  $2.8^{+1.2}_{-1.5}$ &  - &-\\	&&&&&&&&\\	
63 &  12:19:11.67 &   +5:47:28.69 &    - & $45.43 \pm 7.14$&   - & $21.1^{+2.3}_{-2.4}$ & - & - \\
&&&&&&&&\\
64 &  12:19:31.96 &   +5:47:32.34 &   - & $21.56\pm 5.19$ &    - & $4.9^{+1.6}_{-1.9}$ &  - &- \\
&&&&&&&&\\
65 &  12:19:14.69 &   +5:47:38.41 &   - & $10.73 \pm 3.60$&  - &  $2.7^{+0.9}_{-1.0}$ &  - &- \\
&&&&&&&&\\
66 &  12:19:22.55 &   +5:50:16.23 &   $20.94 \pm 5.10$&$8.37 \pm  3.16$ &$3.7^{+1.5}_{-1.9}$ & $2.0^{+0.6}_{-0.8}$ &   A/B & N \\
&&&&&&&&\\
67 &  12:19: 31.25 &   +5:48:17.43 &  - &   $120.79 \pm   11.31 $&  - &   $39.9 ^{+3.4}_{-3.5}$ &    - &- \\
&&&&&&&&\\
68 &  12:19:21.23 &   +5:48:33.69 &   - & $11.08\pm 3.74$ &   - & $5.4^{+1.4}_{-1.6}$ &- &-\\
&&&&&&&&\\
69 &  12:19:36.65 &   +5:49:7.73 &   - & $31.87 \pm 6.00$&- & $7.7^{+1.7}_{-1.7}$ &  -&N \\
&&&&&&&&\\
70 &  12:19:25.91 &   +5:40:15.06 &      - & $6.24 \pm 2.64$ & - & $2.4^{+1.7}_{-2.6}$ & - &-\\
&&&&&&&&\\
71 &  12:19:05.52 &   +5:49:18.92 &    -& $26.88\pm 5.65$ &   - & $15.3^{+3.3}_{-3.4}$ &  - &- \\
&&&&&&&&\\
72 &  12:19:14.44 &   +5:48:48.24 &$10.76 \pm     3.46$ & $8.65 \pm 3.168$& $1.5^{+0.7}_{-0.9}$ & $3.7^{+0.9}_{-1.1}$ &   B &Y \\
&&&&&&&&\\
73 & 12:19:18.67 &   +5:47:43.33 &     $12.61\pm 3.74$&$ 11.26 \pm 3.60$  & $4.4^{+1.7}_{-2.3}$ &$3.8^{+1.1}_{-1.3}$ & B & N \\
&&&&&&&&\\
74 &  12:19:22.66 &   +5:47:22.88 & $13.57   \pm 3.87$ &    -& $4.1^{+1.6}_{-2.2}$ &    - & A/B &- \\

\hline
\end{tabular}
\end{table*}

\begin{table*}
	\centering
	\label{tab:xray2}
	\begin{tabular}{lcccccccr} % four columns, alignment for each
		\hline
		\hline
		Name & RA & Dec & Counts (834) & Counts(9569) & L$_X$  (834) & L$_X$  (9569)  &  HST Field/Giordano & Variability \\
		&&& 0.5-8.0 keV & 0.5-8.0 keV & $10^{38}$ erg s$^{-1}$ & $10^{38}$ erg s$^{-1}$ & & (Y/N) \\
		\hline
	
75 &  12:19:23.64 &   +5:49:39.17 &     - & $10.78  \pm 3.87 $&   - & $3.4^{+2.3}_{-2.5}$ & A &- \\
&&&&&&&&\\
76 &  12:19:24.97 &   +5:47:31.65 & $7.75 \pm 3.16$ & -&$2.6^{+1.4}_{-2.0}$ &    - &   A &- \\
&&&&&&&&\\
77 &  12:19:22.99 &   +5:50:58.48 &$18.29 \pm       4.47$ &     $12.57 \pm 3.74$ & $3.4^{+1.2}_{-1.6}$ &$5.9^{+1.3}_{-1.3}$& A/B & N \\
&&&&&&&&\\
78 &  12:19:12.45 &   +5:49:56.63 & - & $9.59 \pm 3.31$ &  - &    $2.4^{+1.0}_{-1.3}$ &     - &- \\
&&&&&&&&\\

\hline
\hline
\end{tabular}
\end{table*}

% Don't change these lines
\bsp	% typesetting comment
\label{lastpage}
\end{document}